\PassOptionsToPackage{usenames,dvipsnames,svgnames,table}{xcolor}

\documentclass[acmlarge,authorversion]{acmart}

\usepackage{tabularx}
\usepackage{multirow}
\usepackage{makecell}
\usepackage{graphicx}
\usepackage{textcomp}
\usepackage{tikzsymbols}
\usepackage{csquotes}
\usepackage{enumitem}
\usepackage{ragged2e}
\usepackage[capitalise,nameinlink]{cleveref}
\usepackage[subtle]{savetrees}

\microtypecontext{spacing=nonfrench}

\setcopyright{rightsretained}
\acmJournal{DTRAP}
\acmYear{2020}
\acmVolume{1}
\acmNumber{1}
\acmArticle{1}
\acmMonth{1}
\acmPrice{}
\acmDOI{10.1145/3419472}

\newlength{\spc} 
\newcommand{\punct}[1]{%
  \settowidth{\spc}{#1}
  \addtolength{\spc}{-1.8\spc}
  #1
  \hspace*{\spc}
}

\newcommand{\q}[3]{\textit{\enquote{#3} [P#1, #2]}}
\newcommand{\qsimple}[1]{\textit{\enquote*{#1}}}

\newcommand{\personSymbol}{\raisebox{-0.1em}[0pt][0pt]{\large\Strichmaxerl}\hspace{-0.35em}}
\DeclareRobustCommand{\n}[1]{#1\,\personSymbol\nolinebreak\hspace{-0.2em}}

\newcommand{\comprehension}{\textsuperscript{$\star$}}

\acmBadgeR[https://www.acsac.org/2019/]{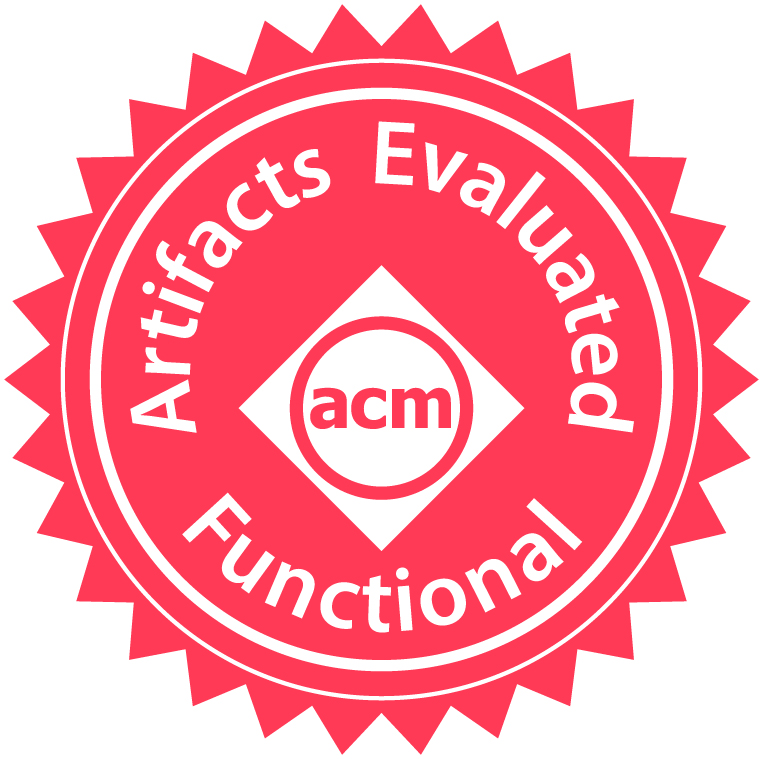}

\begin{document}

\title{Will You Trust This TLS Certificate?}
\subtitle{Perceptions of People Working in IT (Extended Version)}
\titlenote{This is an extended version of the paper published at the \textit{Annual Computer Security Applications Conference (ACSAC) 2019}.}

\author{Martin Ukrop}
\email{mukrop@mail.muni.cz}
\orcid{0000-0001-8110-8926}
\affiliation{%
  \institution{CRoCS, Masaryk University}
  \city{Brno}
  \country{Czech Republic}
}
\author{Lydia Kraus}
\email{lydia.kraus@fi.muni.cz}
\affiliation{%
  \institution{CRoCS, Masaryk University}
  \city{Brno}
  \country{Czech Republic}
}
\author{Vashek Matyas}
\email{matyas@fi.muni.cz}
\affiliation{%
  \institution{CRoCS, Masaryk University}
  \city{Brno}
  \country{Czech Republic}
}

\renewcommand{\shortauthors}{Ukrop and Kraus, et al.}

\begin{abstract}
Flawed TLS certificates are not uncommon on the Internet. While they signal a potential issue, in most cases they have benign causes (e.g., misconfiguration or even deliberate deployment). This adds fuzziness to the decision on whether to trust a connection or not. Little is known about perceptions of flawed certificates by IT professionals, even though their decisions impact high numbers of end users. Moreover, it is unclear how much does the content of error messages and documentation influence these perceptions.

To shed light on these issues, we observed 75 attendees of an industrial IT conference investigating different certificate validation errors. We also analyzed the influence of re-worded error messages and redesigned documentation.
We find that people working in IT have very nuanced opinions with trust decisions being far from binary. The self-signed and the name constrained certificates seem to be over-trusted (the latter also being poorly understood). We show that even small changes in existing error messages can positively influence resource use, comprehension, and trust assessment.
At the end of the paper, we summarize lessons learned from conducting usable security studies with IT professionals.
\end{abstract}



\maketitle

\pagestyle{plain}
\pagestyle{empty}

\section{Introduction}
\label{sec:intro}
Nowadays, communication protected by TLS (formerly SSL) is getting more and more prevalent on the Internet (in May 2019, over 80\% of page loads in Google Chrome were done over HTTPS~\cite{google-https}). For the TLS infrastructure to work, end entities authenticate themselves using X.509 certificates~\cite{rfc-5280}. Certificate validation errors are quite common~\cite{akhawe2013here, acer2017-https-errors}, although an error does not necessarily imply a security incident. For example, getting a self-signed certificate may be either an attack (adversary pretending to be a trusted site) or only a misconfiguration (the local administrator unable or unwilling to obtain a certificate signed by a trusted certificate authority). Imagine the developer connecting to an identity service -- if they accept the self-signed certificate presuming it is just a misconfiguration (when it is not), they will leak authentication credentials of the end users.

Much work has been conducted to understand why system administrators have difficulties in getting TLS deployment and certificate handling right. Usability shortcomings in the TLS deployment process~\cite{krombholz2017have} and related tools~\cite{2018-rsa-ukrop}, as well as other factors beyond the system administrators' control~\cite{dietrich2018investigating}, seem to be common root causes. Furthermore, it has been shown that invalid certificates are sometimes deployed deliberately~\cite{fahl2014eve}.

Nevertheless, most of the previous work on TLS warnings focused on the warning perceptions of end users~\cite{sunshine2009crying, biddle2009browser, Akhawe-2013-warningland, felt2015improving, reeder2018experience}, while the perceptions of certificate flaws by people working in IT seem under-researched. We, therefore, focus on people interacting with software, systems and networks beyond the end user interface (henceforth referred to as \qsimple{IT professionals}\footnote{This includes people responsible for development, testing, deployment and administration alike. Note that we do not limit the sample to any single role since the same errors are encountered by all.}).
IT professionals have a stronger impact on the ecosystem: While the decision of end users to ignore a warning influences only themselves, a similar decision by an IT professional potentially puts numerous end users at risk.

Whenever IT professionals encounter a certificate validation error, we would want them to investigate and draw the right security conclusions. To gain a deeper understanding of whether they can do it, we conducted an empirical study\footnote{Study artifacts are available at \url{https://crocs.fi.muni.cz/papers/dtrap2020} and in the ACM Digital Library.} with 75 attendees of an industrial IT conference. In the test scenario, study participants were confronted with certificate validation errors in a command line interface, asked to investigate the problem and subsequently assess their trust in the certificates (methodology described in \cref{sec:methodology}). Firstly, we look into how IT professionals perceive TLS certificate flaws (\cref{sec:results-opinions}) and examine the resources they use to solve the task (\cref{sec:results-resources}). Secondly, we are interested how much the content of error messages and documentation influences their perceptions. The participants were thus divided into two groups: one with current OpenSSL error messages and documentation and another with re-worded error messages and redesigned documentation. In contrast to the work of Gorski et al.~\cite{new-api-warnings}, we adjust just the content, not the whole concept of the errors and documentation to ensure compatibility.
The comparison of the effects the original and redesigned error messages had are presented in \cref{sec:results-comparison}. After describing the related work (\cref{sec:related-work}) we summarize our experience of usability studies with IT professionals and specifics thereof (\cref{sec:admin-studies}).

\noindent
In summary, our work yields the following main contributions.
\begin{enumerate}
    \item \textit{We provide detailed insights into the perception of different certificate flaws by IT professionals.} 
    We find that their opinions are very nuanced. The trust in flawed certificates is far from binary. Moreover, the self-signed and name constrained certificates seem to be over-trusted (the latter also being poorly understood). The trust in expired certificates heavily depends on the time elapsed from the expiry date.
    \item \textit{We propose and evaluate the influence of redesigned error messages and documentation.}
    We find that the redesigned certificate validation messages and documentation positively influenced resource use, comprehension, and trust assessment.  
\end{enumerate}

\section{Methodology}
\label{sec:methodology}

To gain insights into participants' perception of certificate flaws, we presented them with a scenario-based certificate validation task. Within the task, they successively evaluated multiple server connections with flawed TLS certificates. Participants were divided into two groups with different error messages and documentation.

\begin{figure}[t]
    \centering
    \includegraphics[width=\textwidth]{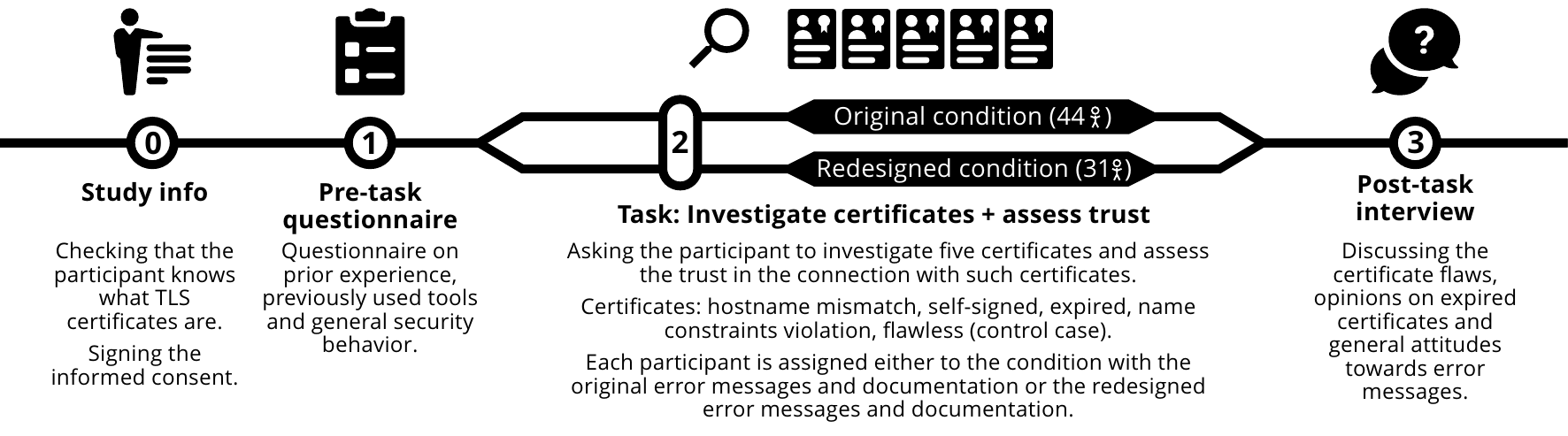}
    \caption{An overview of the experimental procedure. Details are provided in \cref{subsec:methodology-task,subsec:methodology-conditions}. The full questionnaires, task and interview questions can be found in \cref{app:questionnaires}.}
    \label{fig:experiment-overview}
\end{figure}

The experiment overview can be seen in \cref{fig:experiment-overview}.
The study had a mixed design: The participants were validating five different certificates (the within-subjects factor) across two different conditions (original or redesigned documentation; the between-subjects factor).

\subsection{Procedure}
\label{subsec:methodology-task}
The study consisted of three parts: 1) an introductory questionnaire on previous experience, 2) a certificate validation task with trustworthiness assessment and 3) an exit semi-structured interview followed by an educational debriefing. All parts of the experiment were in English, except for the interview, which could optionally be in Czech. The study had a mixed design: The participants were validating five different certificates (the within-subjects factor) across two different conditions (original or redesigned documentation; the between-subjects factor).
The order of the certificates was randomized and participants were randomly assigned to one of the between-subjects conditions.

\subsubsection{Initial Questionnaire}

At the beginning, we inquired into the previous experience of the participants: the years of employment, formal education, self-reported knowledge of computer security and X.509 certificates in particular (5-point Likert scale from \textit{poor} to \textit{excellent} for the last two questions). To make sure the sample uses relevant tools, we asked about previous experience with OpenSSL and other certificate manipulation utilities. Afterward, they received 16 questions of the Security Behavior Intentions Scale (SeBIS)~\cite{egelman2015scaling}. The full questionnaire can be found in \cref{app:questionnaires}.

\subsubsection{Certificate Validation Task}
\label{subsubsec:methodology-task}

Participants were asked to evaluate their trust in five different server certificates (in a randomized order). The scenario stated they were implementing a patch for the conference website to support registration using external identity providers (Microsoft, Fedora Project, GitHub, Google and Facebook). It was said that they had started by writing a small program establishing a TLS connection to the providers' servers. However, since some certificates caused validation errors, the participants were asked to investigate these and evaluate the trustworthiness of the connections.

When running the given program for each identity provider in the command line interface, a short certificate validation message was displayed. Apart from that, the certificate was \enquote{downloaded} and saved next to the program binary to allow for further (manual) inspection. The participants were not limited in their investigation -- they could inspect the certificate, browse the Internet or use local sources for as long as they wanted. For convenience, the printed version of the task contained a command to view the downloaded certificate. The complete task can be found in \cref{app:questionnaires}.

Below, we discuss the five certificate issues used in our experiment. We included three cases commonly occurring in the wild, one extension that would allow for a fine-grained control in the ecosystem (yet is still rarely used) and a control case.
\begin{itemize}
    \item \textbf{Hostname mismatch.} The certificate that was presented by \textit{oauth.facebook.com} had a slightly different subject (\textit{oauth.facesbook.com}, an extra \enquote*{s} in the name). Browser-based measurements show roughly 10\%~\cite{acer2017-https-errors} to 18\%~\cite{akhawe2013here} of the certificate errors due to a name mismatch.
    \item \textbf{Self-signed.} The certificate for \textit{id.fedoraproject.com} was signed by its own key (it did not chain up to a trusted root). 
    Such errors constitute roughly 1\% of browser certificate errors~\cite{acer2017-https-errors}, but their prevalence in TLS scans can be as high as 25\%~\cite{holz2011ssl} or 88\%~\cite{silent-majority}.
    \item \textbf{Expired.} The certificate was issued for \textit{login.microsoft.com} and its validity ended seven days before the experiment. 
    Expired certificates cause from 4\%~\cite{acer2017-https-errors} to 7\%~\cite{akhawe2013here} of browser certificate errors and are present at about 18\% of the sites~\cite{holz2011ssl}.
    \item \textbf{Name constraints.} The chain for \textit{accounts.google.com} contained an intermediate authority that was constrained (by the root authority) to issue certificates in the domain subtree \textit{*.api.google.com}. Thus, this violated name constraint invalidated the certificate chain. The name constraints extension was standardized back in 2002 in RFC 3280~\cite{rfc-3280} and the CA/Browser forum guidelines allow for its use on public websites since June 2012~\cite{cabforum-br}. Despite its potential to prevent certificate misissuance, only 0.001\% of the unexpired certificates in May 2019 used this extension~\cite{censys}.
    \item \textbf{OK.} The certificate for \textit{auth.github.com} had no validation issues.
    It was included as the control case.
\end{itemize}
All the certificates (apart from the self-signed), were chained up to a self-signed certificate for the experiment root CA. The participants were explicitly instructed (in the task specification) to trust this root.

After each case, the participant was asked to assess the trustworthiness of the connection with such a certificate. A 7-point scale was provided. It spanned from \qsimple{Trust 6/6: I'm totally satisfied. If it were my bank's website, I would log in without worries.} down to \qsimple{Trust 0/6: Outright untrustworthy. It is not safe to browse or to trust any information there.} The full scale can be found in \cref{app:questionnaires}. 
This Likert scale for trust was adopted from our related prior experiment on certificate trust evaluation by university computer science students. That original experiment had the 4-point scale (only the labeled items). However, it turned out not to be fine-grained enough: the case differences were small and the participants mentioned they sometimes wanted to choose \enquote{something in between.} We had bound the original trust levels to specific behaviors and adding additional labels to levels turned out to be difficult (e.g., finding a behavioral equivalent for something between \enquote{not visiting a site} and \enquote{visiting, without entering information}). Subsequently, we decided to provide the \enquote{in-between options} without labels and let the participants share their interpretations in the post-task interview.

\subsubsection{Post-task Interview}

After the certificate validation, a short semi-structured 
interview took place. First and foremost, the participants were asked to explain in their own words the problem with each certificate.
Retrospective evaluations help gain valuable insights into participants' reasoning, especially in the context of usability studies with advanced users. Due to the task complexity, this approach was more suitable than methods that foster immediate reflection, such as think-aloud protocols \cite{naiakshina2017developers}. After discussing the individual certificate issues, expired certificates were talked over in detail. The full list of base questions guiding the interview can be found in \cref{app:questionnaires}.
To educate the IT community, the experimenter clarified any misconceptions or misunderstandings in the debriefing after the participants described their view. 

\subsection{Experimental Conditions}
\label{subsec:methodology-conditions}

We deployed two experimental conditions differing in the error messages and documentation with each participant randomly assigned to one condition. The fact that they may have alternative error messages was only disclosed in the debriefing after completing the task.

The first condition, henceforth referred to as the \qsimple{original}, used current error messages from OpenSSL 1.1.0g-fips (build from November 2\textsuperscript{nd}, 2017) with the corresponding documentation (manual pages and resources available on the Internet). We decided to use OpenSSL as a base condition since it is one of the most popular crypto libraries for generating keys/certificates \cite{nemec2017measuring}.
The original error messages contain several descriptive words and a textual error code that can be looked up in the documentation. The documentation is available in the form of manual pages and the online OpenSSL Wiki~\cite{openssl-wiki}. However, the descriptions in the documentation contain only little extra information compared to the actual error message.

The second, \qsimple{redesigned}, condition had re-worded error messages with a link to the website containing our custom-written documentation. We expected the documentation might have room for improvement due to related research on poor usability of OpenSSL~\cite{2018-rsa-ukrop}. To ensure compatibility, we decided not to design a completely novel system of errors (compared to the work by Gorski et al.~\cite{new-api-warnings}).

The first draft of the new error messages and documentation was done by the principal researcher synthesizing personal experience from open-source development of certificate-validating software, existing documentation and designs by multiple students (asked to create usable documentation for the chosen certificate validation error). 
Subsequently, the draft underwent multiple iterations between the researcher and a usable security expert with experience from both industry and academia. Overall, the design process followed the principles of warning design proposed by Bauer et al.~\cite{bauer2013-warning-guidelines}.

The error messages were re-worded to explain the cause in more detail, but to still fit into a single line (including the link to the new documentation page). We decided to keep the text codes for easy reference and reasonable compatibility. The documentation for each case has a consistent structure of four subsections: 1) summarizing the problem; 2) explaining what happened; 3) giving the security perspective (explaining the risks) and 4) proposing what steps could be taken next. The re-worded error messages and the full documentation can be seen in \cref{app:documentation}.

\subsection{Pilot Testing, Recruitment and Study Setup}
\label{subsec:methodology-setting-recruitment}

We pilot-tested the study in two iterations. Firstly, we had three participants, who were experienced with usability studies. We performed a combination of the think-aloud protocol and post-testing discussion. This iteration was used to adjust the task, questionnaire and interview content and get approximate timing information. Secondly, we recruited four participants who undertook the experiment in full setup (including the recordings and measurements). Testing had shown no issues in the experiment design, but uncovered some smaller technical glitches and task ambiguities.

The final study took place in 2018 at the DevConf.CZ conference\punct{.}\footnote{DevConf.CZ is an international community conference for developers, admins, devops engineers, testers and others interested in open source technologies such as Linux, middleware, virtualization or storage. The conference is organized by Red Hat Czech and had approximately 1700 attendees in 2018.}
We had set up a booth offering participation in research regarding the usability of developer tools. The booth featured three Linux workstations with (virtualized) Fedora 27 (a standard workstation flavor) connected to the Internet. The operating system logged executed commands and browsed websites as well as recorded the screen.

The participants were not primed to expect security tasks, but before accepting them, we made sure that they knew what TLS certificates were. We decided not to limit our sample to security professionals, as it turns out that the majority of developers work with security features from time to time~\cite{gorski2017-security-apis}. The whole experiment took about 30--40 minutes per participant (pre-task questionnaire, task, post-task interview). 
For completing the task, we gave participants merchandise of the organizing company (i.e., there was no monetary compensation).

All the participants were briefed about the extent of processed personal information. Participation was voluntary and the audio recording of the interview was optional. All the data was collected pseudonymously. As stated in the informed consent form, the subset of the data (excluding the screen and interview recordings) were anonymized and are published as an open dataset. The study design has been approved by our research ethics committee.

\subsection{Participants}
\label{subsec:methodology-participants}

From the 78 recruited participants\punct{,}\footnote{From now on, we use the symbol \personSymbol\ to denote the participants.} \n{3} were omitted due to not completing all the sub-tasks, resulting in the base sample count of \n{75}. 
The participants were employed in the IT sector on average for $9.55 \pm 7.06$\footnote{The plus-minus symbol ($\pm$) is used to denote standard deviation.} years ($median\;\;8$). In more detail, 16\% of the participants (\n{12}) were employed for 0--2 years, 25\% (\n{19}) for 3--5 years, 16\% (\n{12}) for 6--8 years, 11\% (\n{8}) for 9--11 years and 32\% (\n{24}) for 12 or more years (yet nobody for more than 30 years).

The formal education levels were also diverse: 33\% of the participants (\n{25}) did not have a formal degree in computer science or engineering, 25\% (\n{19}) had a bachelor's degree, 38\% (\n{29}) a master's and 3\% (\n{2}) a Ph.D.
Participants' self-reported security knowledge was \qsimple{good} on average ($mean\;\;1.91 \pm 1.02$, $median\;\;2$), their experience with X.509 certificates was \qsimple{fair} on average ($mean\;\;1.00 \pm 1.07$, $median\;\;1$).
Almost all of the participants (91\%, \n{68}) had used OpenSSL before, followed by Network Security Services (25\%, \n{19}), Java Keytool (25\%, \n{19}), GnuTLS (19\%, \n{14}), and Windows Certutil (12\%, \n{9}). About a third had used the Let's Encrypt Certbot before (36\%, \n{27}).

To estimate the security behavior of our participants, we used the  \textit{Security Behaviors Intentions Scale} (SeBIS)~\cite{egelman2015scaling}.
Our sample tends to have stronger intentions towards secure behavior than the general population~\cite{egelman2015scaling} -- the individual subscale scores on average 0.1--1.3 points higher with the highest difference for the device securement subscale.

The vast majority consented to the audio recording of the interview, forming the sample size for the qualitative analyses (89\%, \n{46} in English, \n{21} in Czech). 59\% of the participants (\n{44}) were (at random) assigned to the original condition with the remaining 41\% (\n{31}) assigned to the redesigned conditions. Checking for the differences across conditions, we see no significant differences with respect to self-reported previous experience, IT education, IT employment or used tools suggesting that the samples are comparable.

To check for potential biasing effects, we asked if the participants had taken part in an experiment conducted at the same conference in the previous year~\cite{2018-rsa-ukrop}.
Only a minority of the participants did (13\%, \n{10}). Omitting these participants did not alter the conclusions drawn from the results (we compared all the results for the whole sample and the sample omitting these participants). We, therefore, present the results of the full sample.

Lastly, we check for potential differences between experimental conditions (original vs. redesigned errors). Looking at previous knowledge, we found no significant differences in the number of years spent in IT, formal education in computer science, self-reported knowledge of computer security or X.509 certificates (all assessed by the Mann-Whitney U tests). Previously used tools to manipulate X.509 certificates also have no significant differences between conditions (individual tool differences tested by Fisher's exact test, the number of previously used tools by the Mann-Whitney U test). SeBIS questions (both individual and subscale means) also don't exhibit any significant differences (Mann-Whitney U test).

\subsection{Data Collection and Analysis}
\label{subsec:methodology-analysis}

First of all, data from all sources was matched based on timestamps and each participant had a pseudonym assigned. Timestamps were used to compute the time spent on the individual tasks. The browser history was utilized to identify used resources.

We performed a qualitative descriptive analysis of the post-task interview~\cite{sandelowski2000whatever}. After transcribing the interviews, two researchers independently familiarized themselves with the data.
The data was then processed using open coding~\cite{saldana2015coding} (the researchers independently looked for reoccurring themes and assigned codes to text passages). 
The analysis was open but broadly framed by the participants' comprehension of the encountered validation messages, their opinions and actions.
After the open coding, the researchers discussed the created codes and consolidated a common codebook. The transcripts were then re-coded using the shared codebook. 
The results reported in \cref{sec:results-opinions,sec:results-comparison} are based on this coding round.

To ensure analysis reliability, a third independent coder coded half of the English and half of the Czech transcripts using the shared codebook.
Interrater agreement between the original coders and the independent coder showed to be \textit{substantial} (according to Landis and Koch~\cite{landis1977measurement}) for both English (Cohen's $\kappa = 0.69$, $p <0.001$) and Czech (Cohen's $\kappa = 0.63$, $p <0.001$).
All coders used all codes, indicating that the defined codes were actually present in the interviews and that the codebook was defined well. 
Some deviations that appeared in the coding stemmed from the fact that the independent coder was slightly more conservative in assigning codes compared to the original coders. 

Supplementary materials, including the experimental setup (the complete virtual machine) and the anonymized dataset are available at \url{https://crocs.fi.muni.cz/papers/dtrap2020} and in the ACM Digital Library.

\subsection{Study Limitations}
\label{subsec:methodology-limitations}

As is the case with every study, various limitations may diminish the applicability of results. First, to ensure that participants would behave as usual, we designed a realistic and appealing task (at a conference on open-source technologies, we had them \enquote{develop} a patch for the registration system). We limited neither time nor resources allowing the participants to behave as they would in reality. 
Realistic and well-known entities and hostnames were chosen for the certificates. Although different hostnames may have slightly different reception, we preferred this to the ecologically less valid option of solving five separate cases with the same hostname.

The second great concern is the sample bias -- we recruited attendees of a single industrial conference. Nevertheless, we think our sample reflects the wider population of people working in IT well enough. 
Even though the conference was biased towards open-source and Linux technologies, a 2018 Developer Survey by Stack Overflow~\cite{2018-so-dev-survey} estimates that almost half of the developers contribute to open source and Linux is the most developed-for platform.
Compared to the mentioned survey, our sample follows the general trends for professional experience and education but has a slightly higher mean (peaking at 3--5 years working in IT instead of 0--2 and at the master's degree instead of the bachelor's).
This means the reported comprehension may be a bit higher for our sample compared to the wider population (as our sample is a bit more experienced and educated).

Thirdly, participants' behavior may have been primed towards security by the context or parts of the questionnaire. However, we were cautious not to mention security when advertising the study and tried to recruit all attendees passing by regardless of their skill.

Lastly, we tried to mitigate multiple response biases. To combat the question order bias, we randomized the order of questions in the initial questionnaire and the order of evaluated certificate cases both in the task and the trust scale. Furthermore, to work against the bias of the response order, we inverted all the Likert scales in both the questionnaire and trust assessment for half of the participants. To lower the observer effect (participants behaving differently when being watched), the interviewers left the participants alone when completing the tasks (but were still available in case the participant wanted to consult something). 
Some participants might have been more cautious than usual due to the social desirability bias -- to account for this, we see the obtained trust evaluations as a lower bound.

\section{Perception of Certificate Flaws}
\label{sec:results-opinions}

In this section, we present comprehension and perceived trustworthiness of tested certificate flaws, together with the reasoning that participants provided in the exit interview. Comprehension is based on qualitative analysis of the post-task interview (\n{67}). Trust assessment and answers to the structured interview questions are available for everybody (\n{75}). The codes resulting from the qualitative investigations can be seen in \cref{tab:interview-codes,tab:interview-codes2} along with their frequencies, simplified definitions and representative quotes. The section ends with a cross-case comparison and investigation into the influence of previous knowledge.

\begin{table}
\centering
\caption{Overview of the comprehension and reasoning codes occurring in at least 10\% (\n{7}) of the interviews (\n{67} in total, \n{39} in the original (\textsc{o.}) and \n{28} in the redesigned (\textsc{r.}) condition). Codes labeled with an asterisk (\comprehension) indicate case comprehension.}
\label{tab:interview-codes}
\rowcolors{2}{gray!20}{white}
\setlength{\tabcolsep}{3pt}
\begin{tabularx}{\textwidth}{>{\cellcolor{white}\bf}p{0.6cm}@{}>{\sc}l<{\hspace{0.15em}}|>{\hspace{0.2em}}r@{\hspace{0.5em}}>{(}c<{)\hspace{0.25em}}>{\RaggedRight\vspace{0.2em}}m{3.5cm}@{\hspace{0.5em}}>{\RaggedRight\arraybackslash}m{7cm}}
\rowcolor{white}
\multicolumn{2}{l|}{\textsc{\textbf{Case} / Code}} & \personSymbol & \textsc{o.+r.} & \textsc{Code definition} & \textsc{Representative quote} \\ \hline

    & BadName\comprehension & \n{50} & 29+21 & Cert. subject and server name do not match. &
        \q{39}{original}{The last one server, Facebook, [the certificate] was issued for a~different hostname.} \\
    & NameCheck & \n{27} & 16+11 & Mentioning the exact difference in the names. &
        \q{57}{original}{[...] because it is not Facebook, it is Facesbook or something like that.} \\
    & Attack & \n{22} & 12+10 & Connection may be attacked. &
        \q{76}{original}{It can be some phishing site or something like this.} \\
\multirow{-7.5}{4em}{\rotatebox[origin=c]{90}{Hostname mismatch}}
    & Mistake & \n{8} & 3+5 & It can be only a mistake or server misconfiguration. &
        \q{63}{redesigned}{And in this case -- it's a different domain, but I'd say it's some kind of~typo or something like that.} \\
    \hline

    & ByItself\comprehension & \n{50} & 30+20 & The certificate is signed by~itself/self-signed. &
        \q{15}{original}{That it is not signed by the other authority, but it's signed by itself.} \\
    & NoCA\comprehension & \n{28} & 16+12 & No CA was involved in~issuing~the~certificate. &
        \q{20}{original}{It was signed by local server for which it was generated. It was not signed by official authority.} \\
    & AnyoneCan\comprehension & \n{21} & 12+9 & Anyone can create such~a~certificate. &
        \q{78}{original}{Self-signed certificate? Anyone can create self-signed certificates.} \\
    & IfExpected & \n{10} & 6+4 & It is OK if such certificate is~known/expected. &
        \q{09}{original}{If I knew that the certificate should be self-signed, I could consider it trustworthy.} \\
    & Internal & \n{10} & 7+3 & They are used for testing or internal purposes. &
        \q{11}{original}{[...] and it's usually used either by internally or for testing purposes. It~shouldn't be used publicly.} \\
\multirow{-13}{*}{\rotatebox[origin=c]{90}{Self-signed}}
    & Attack & \n{8} & 2+6 & Connection may be attacked. &
        \q{66}{original}{[...] because that can be any hacker, [they] can phish.} \\
    \hline
    
    & NoLonger\comprehension & \n{62} & 34+28 & The certificate has expired. &
        \q{30}{original}{Microsoft certificate has expired, it's out of date.} \\
    & Mistake & \n{27} & 15+12 & It can be only a mistake or server misconfiguration. &
        \q{10}{original}{[...] it could be just forgotten and they're about to do it, they're about to renew it or something.} \\
    & Common & \n{18} & 12+6 & Expired certificates are common/happen. &
        \q{01}{redesigned}{Ah, right, so, expired certificates are pretty common, so from what I~can see [...]} \\
    & OKBefore\comprehension & \n{14} & 4+10 & The certificate was OK in~the~past. &
        \q{18}{redesigned}{Yeah, the Microsoft one is expired. So it was valid in the past, and I~looked at the date [...]} \\
    & Reputation & \n{13} & 7+6 & Taking into account the~subject~of the certificate. &
        \q{62}{original}{If it's like a small businesses from my local neighborhood, I would probably trust them.} \\
\multirow{-14}{*}{\rotatebox[origin=c]{90}{Expired}}
    & Attack & \n{8} & 4+4 & Connection may be attacked. &
        \q{37}{original}{[maybe] the attacker has stolen a certificate which was previously valid and has been revoked [...]} \\ 
\end{tabularx}
\end{table}

\begin{table}
\centering
\caption{Overview of the comprehension and reasoning codes occurring in at least 10\% (\n{7}) of the interviews (\n{67} in total, \n{39} in the original (\textsc{o.}) and \n{28} in the redesigned (\textsc{r.}) condition). Codes labeled with an asterisk (\comprehension) indicate case comprehension.}
\label{tab:interview-codes2}
\rowcolors{2}{gray!20}{white}
\setlength{\tabcolsep}{3pt}
\begin{tabularx}{\textwidth}{>{\cellcolor{white}\bf}p{0.6cm}@{}>{\sc}l<{\hspace{0.15em}}|>{\hspace{0.2em}}r@{\hspace{0.5em}}>{(}c<{)\hspace{0.25em}}>{\RaggedRight\vspace{0.2em}}m{3.5cm}@{\hspace{0.5em}}>{\RaggedRight\arraybackslash}m{7cm}}
\rowcolor{white}
\multicolumn{2}{l|}{\textsc{\textbf{Case} / Code}} & \personSymbol & \textsc{o.+r.} & \textsc{Code definition} & \textsc{Representative quote} \\ \hline

    & Constraint\comprehension & \n{25} & 12+13 & The name of the endpoint certificate is constrained. &
        \q{39}{original}{I understood that there is some chain and a certain point in chain is~restricting the hostname to ...} \\
    & Wrong & \n{19} & 12+7 & Giving reasoning that~is~wrong. &
        \q{10}{original}{I find out that one of the authorities was listed as false, but the other two were fine.} \\
    & NotKnow & \n{14} & 11+3 & I do not understand it. &
        \q{62}{original}{I don't really understand the whole thing.} \\
    & Attack & \n{10} & 3+7 & Connection may be attacked. &
        \q{26}{redesigned}{[I'd] let Google know that they have a~rogue admin...} \\
    & CAProblem\comprehension & \n{10} & 2+8 & The intermediate CA was not~allowed to issue this. &
        \q{26}{redesigned}{[...] CA has explicitly said \enquote{I am not allowed to sign this, you should not trust this.}} \\
    & CAConstr\comprehension & \n{9} & 2+7 & The constraints are set by~the~CA. &
        \q{18}{redesigned}{The certificate authority up the chain specifies that only domains with \enquote{api.google.com} are valid.} \\
    & Mistake & \n{7} & 3+4 & It can be only a mistake or server misconfiguration. &
        \q{19}{original}{It seemed like it was just an innocent misconfiguration of the kind that happens all the time.} \\
\multirow{-20}{4em}{\rotatebox[origin=c]{90}{Name constraints violation}}
    & NoInfo & \n{7} & 5+2 & Finding more information on~the matter is difficult. &
        \q{68}{original}{For this one I really try to find some documentation, but there was no documentation on this.} \\
    \hline
        
    & NoIssue\comprehension & \n{61} & 34+27 & There is no problem. &
        \q{22}{original}{There wasn't a problem, it was good, OK.} \\
    & ExtraCheck & \n{13} & 5+8 & Doing extra manual checks\newline on the certificate. &
        \q{13}{redesigned}{I think it was safe, but I looked into the cert and I couldn't find anything wrong, so I would trust it...} \\
\multirow{-6}{*}{\rotatebox[origin=c]{90}{OK (flawless)}}
    & BugFree & \n{12} & 7+5 & The program is trusted to do the verification correctly. &
        \q{77}{redesigned}{[...] everything looked fine and I thought: \enquote{Well, if the testing tool is~good, I'll trust that.}}
\end{tabularx}
\end{table}

\subsection{Hostname Mismatch Case}

The hostname mismatch flaw was comprehended quite well -- a majority of the participants mentioned the core of the issue was the server hostname not matching the name provided in the certificate (code \textsc{BadName}, \n{50}, see \cref{tab:interview-codes}). Many participants explicitly mentioned the extra letter \enquote*{s} in the certificate name (\textsc{NameCheck}, \n{27}), hinting at the fact that they looked at the certificate to investigate the issue. The prevalent opinion regarding the cause of the error was that it was an attack of some sort (\textsc{Attack}, \n{22}), but a few participants explicitly mentioned it could be only a mistake or a typo (\textsc{Mistake}, \n{8}).

\begin{figure}
    \centering
    \includegraphics[width=\textwidth]{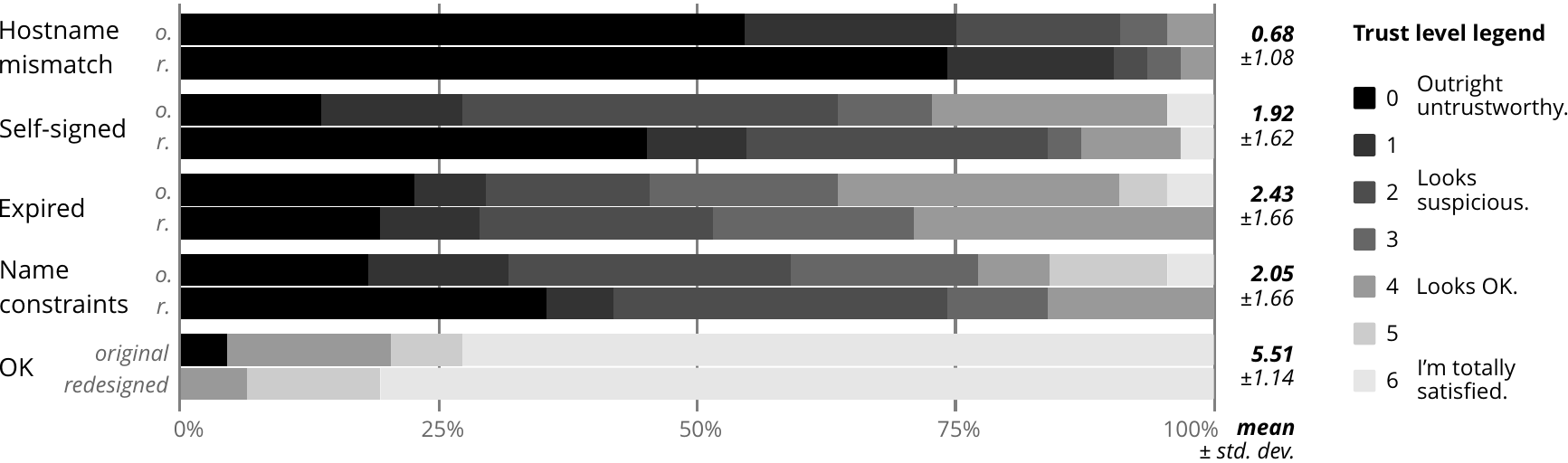}
    \caption{Comparison of trust assessment among certificate cases split by condition.
    Bars are normalized due to the different number of participants in each condition (\n{44} for the original condition, \n{31} for the redesigned one).
    }
    \label{fig:trust-comparison}
\end{figure}

The connection was on average assessed as \qsimple{being outright untrustworthy} ($mean\;\;0.68 \pm 1.08$, $median\;\;0$, see \cref{fig:trust-comparison} for detailed information and \cref{app:questionnaires} for the whole scale). We find the low trust in this case unsurprising as the server with mismatched second-level domain name gives almost no identity guarantees.

\subsection{Self-signed Case}
\label{subsec:self-signed}

The self-signed case is also dominated by codes indicating comprehension: Mentioning that the certificate is self-signed (\textsc{ByItself}, \n{50}), that no CA was involved in its issuance (\textsc{NoCA}, \n{28}) and that literally anyone (including the participants themselves) can issue such a certificate (\textsc{AnyoneCan}, \n{21}). Some participants mentioned that they would find self-signed certificates acceptable in places where they were told to expect them (\textsc{IfExpected}, \n{10}). The self-signed certificates were multiple times said to be OK for testing or internal purposes (\textsc{Internal}, \n{10}). A few participants mentioned the possibility of an attack when encountering self-signed certificates (\textsc{Attack}, \n{8}).

The connection with a self-signed certificate was perceived as \qsimple{looking suspicious} ($mean\;\;1.92 \pm 1.62$, $median\;\;2$, see \cref{fig:trust-comparison}). The rating similar to the expired case (see \cref{subsec:results-expired}) is surprising, given that the expired certificate had at least provided an identity guarantee in the past, whereas the self-signed certificate provides literally no identity guarantees and never did. 

\subsection{Expired Case}
\label{subsec:results-expired}

When participants described their perception of the problem, most of them seemed to comprehend the source of the flaw -- the certificate being no longer valid (\textsc{NoLonger}, \n{62}) or having been valid in the past but not now (\textsc{OkBefore}, \n{14}). The prevalence of the first code is unsurprising as it is literally the content of the displayed error message.

Several considerations seemed to play a role in opinion forming: Many thought expired certificates can be grounded in a misconfiguration, negligence or operator error on the server side, and thus the flaw is non-malicious (\textsc{Mistake}, \n{27}). A considerably lower number of participants mentioned that the connection might be attacked or an old certificate may be abused (\textsc{Attack}, \n{8}).

Next, multiple participants expressed the opinion that expired certificates are quite common (\textsc{Common}, \n{18}). The last notable mention is the participants binding the trust in expired certificates to the subjects using them (\textsc{Reputation}, \n{13}) -- seeing an expired certificate as more acceptable for local business than for Microsoft (the specific comparison was drawn because the expired certificate used in the task was issued for \textit{login.microsoft.com}).

Participants rated the trustworthiness of the connection with the server certificate expired one week ago as rather low, close to a \qsimple{looks suspicious} point on our scale ($mean\;\;2.43 \pm 1.66$, $median\;\;3$, see \cref{fig:trust-comparison}). Given that the provided certificate does not guarantee the identity information anymore, it would have been unsurprising to us if the trustworthiness assessment went even lower.

\begin{figure}
    \centering
    \includegraphics[width=\textwidth]{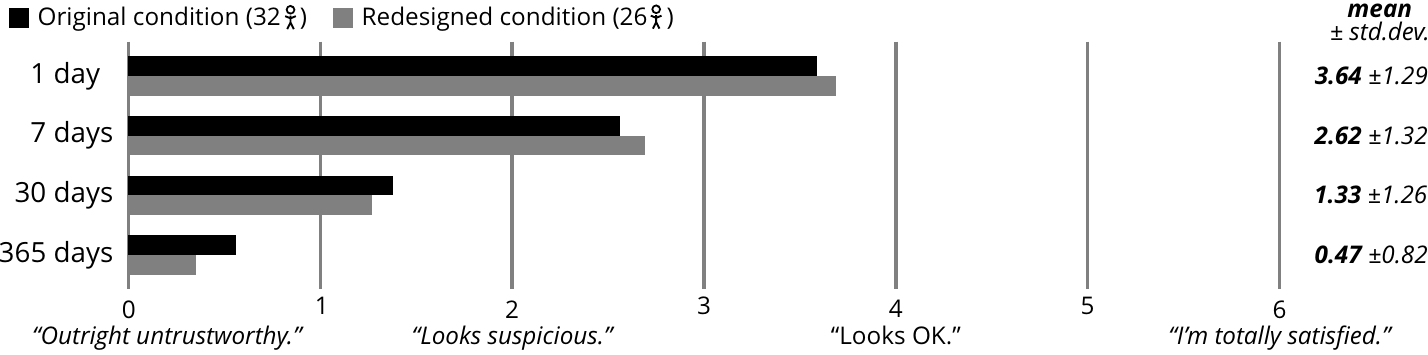}
    \caption{Mean trust in the certificates depending on how long it has been expired, compared across conditions. Only participants with different ratings for durations are included.}
    \label{fig:expired-trust}
\end{figure}

From our own experience as programmers and IT security educators, we further suspected the trust might be influenced by the expiry duration. Subsequently, we had prepared several follow-up questions for the expired case in the exit interview.
Fair enough, the majority of participants (77\%, \n{58}) reported that the rating would be different for shorter/longer periods past the expiration date. However, only 65\% of these (\n{36}) actually checked the expiry date in the given task (an example for the not uncommon difference between self-reported intention and behavior). When subsequently asked to rate trust in the connection with the server certificate expired a day/week/month/year ago, the mean trust spanned from \qsimple{looking OK} ($mean\;\;3.64 \pm 1.30$, $median\;\;4$) for a day-expired certificate to almost \qsimple{outright untrustworthy} ($mean\;\;0.47 \pm 0.82$, $median\;\; 0$) for the certificate expired a year ago. The detailed gradual decline of trust can be seen in \cref{fig:expired-trust}. The results indicate significant differences in trust among the differently old expired certificates (Friedman ANOVA, $\chi^2(3)=141.58$, $p<0.001$). Almost all pairwise comparisons (except for month/year) were significant (Dunn-Bonferroni post-hoc analysis, $p<0.005$). The effect size was the largest for the day/year comparison ($r=0.22$), followed by day/month ($r=0.16$) and week/year ($r=0.15$). This confirms that expiry is not a binary feature and people's trust significantly depends on the time elapsed from the expiry date. 

When asked about the relation of the expiry date and the certificate revocation, only 24\% of the participants (\n{18}) knew that expired certificates are no longer included in the certificate revocation lists (CRL). Of the \n{57} who did not know this, \n{25} indicated that knowing it would lower their trust rating. That suggests that insufficient knowledge of the ecosystem may cause over-trusting the certificates.

\subsection{Name Constraints Case}

The participants' problem descriptions in this case suggest only a limited comprehension of the flaw and its security implications. Less than half of the participants mentioned that the domain name in the endpoint certificate is constrained or that there is a blacklist/whitelist for these names (\textsc{Constraint}, \n{25}). Even fewer people said that the CA was not allowed to issue the endpoint certificate (\textsc{CAProblem}, \n{10}) or that it is the CA who is constraining the name (\textsc{CAConstr}, \n{9}). More participants were able to describe the problem in the redesigned case (see details in \cref{subsec:conditions-comprehension}).

The name constraint was often surrounded by explicit statements of not understanding the problem (\textsc{NotKnow}, \n{14}) or difficulty to obtain further information (\textsc{NoInfo}, \n{7}). This was expected, as name constraints are little used (and thus little known) and the error message in the original condition only says \qsimple{permitted subtree violation} without further details. The notion of name constraints being the most complicated case is further supported by the opinion of the participants themselves -- when asked in the post-task interview what task they saw as the most difficult to understand, almost all of them (90\%, \n{61}) indicated the name constraints case.

A high count of people provided wrong reasoning (\textsc{Wrong}, \n{19}) for the cause of the problem. A few participants (\n{5}) considered the case to be the same as the hostname mismatch case. The others blamed other (often unrelated and flawless) parts of the certificates: the basic constraints extension (\qsimple{the CA is set to false}), the key purpose extension (\qsimple{the purpose is wrong}), the name format (\qsimple{there are wildcards}, \qsimple{the third-level domain is missing}) or other certificates in the chain (\qsimple{the chain integrity is broken}, \qsimple{the intermediate CA certificate is missing}).
Note that misconceptions and the lack of understanding were more often mentioned in the original condition (see \cref{subsec:conditions-comprehension}).
Of the few opinions on the causes of the validation error, more participants saw it as a potential attack (\textsc{Attack}, \n{10}) than a mistake or a misconfiguration (\textsc{Mistake}, \n{7}).

The average trust into the connection was \qsimple{looking suspicious} ($mean\;\;2.05 \pm 1.66$, $median\;\;2$, see \cref{fig:trust-comparison}). We see it as surprising that the name constraints error is rated similarly to the expired case, given that its security implications are potentially more severe (the intermediate authority being corrupted).
The described misunderstandings further suggest the error message does not sufficiently pinpoint the relevant issue in the certificate structure.

\subsection{OK Case (Flawless)}

Concerning the control case, the majority of participants stated there was nothing wrong with the certificate (\textsc{NoIssue}, \n{61}). Some participants did an extra check manually (\textsc{ExtraCheck}, \n{13}), though only a few of them (\n{3}) hinted this behavior was influenced by the experimental environment. Many mentioned that their opinion and trust is based on the fact that they were instructed to presume that the used program is bug-free (\textsc{BugFree}, \n{12}).

Regarding the perceived trust in connections where the server is using such a certificate, participants were on average close to \qsimple{being totally satisfied} with the certificate that validated without errors ($mean\;\;5.51 \pm 1.14$, $median\;\;6$, see \cref{fig:trust-comparison}). We do not find this result surprising, given that the certificate was flawless.

\subsection{Case Comprehension and Trust}
\label{subsec:results-opinions-comparing}

\begin{figure}
    \centering
    \includegraphics[width=\textwidth]{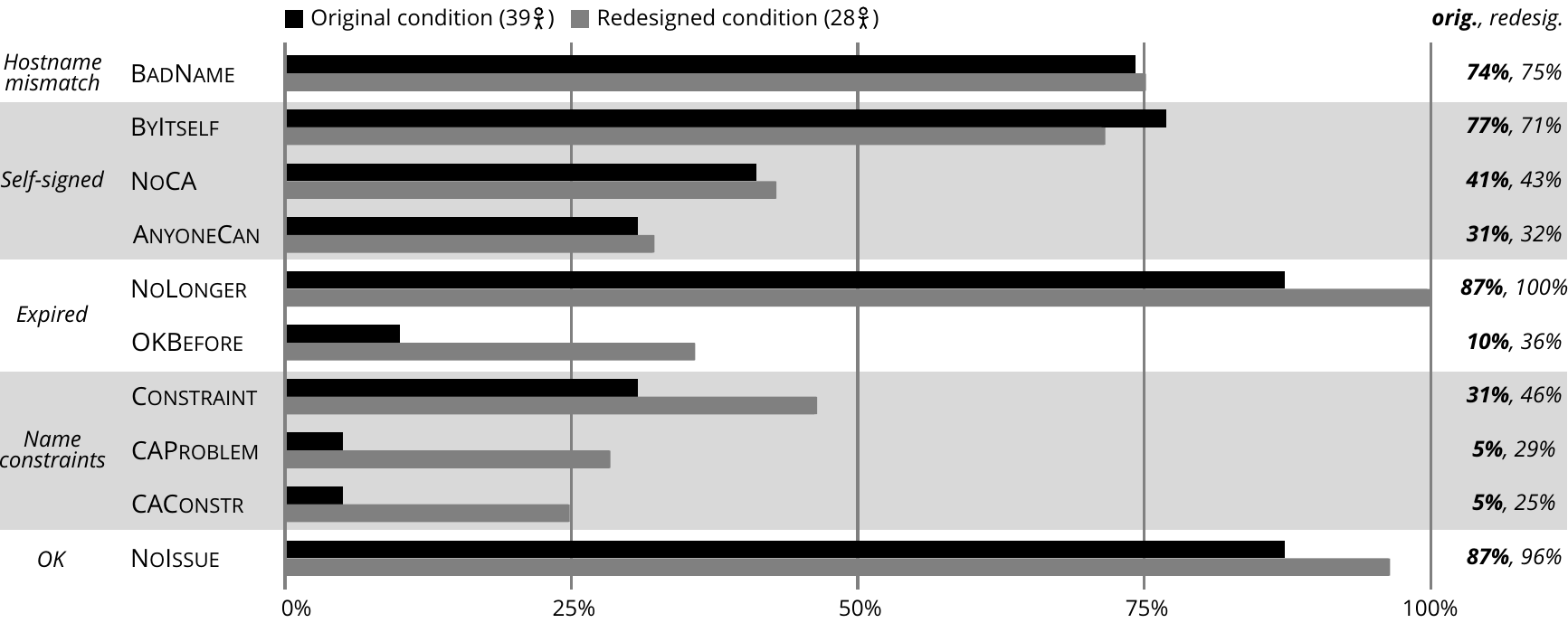}
    \caption{An overview of codes indicating case comprehension for all cases, compared across conditions. For more details on the codes, see \cref{tab:interview-codes,tab:interview-codes2}.}
    \label{fig:understanding-comparison}
\end{figure}

Let us now compare and contrast the individual cases. We reason about case comprehension by analyzing frequently occurring interview codes describing the problem, see \cref{fig:understanding-comparison}. Note that this measure is only approximative -- (not) mentioning the particular things does not necessarily mean (not) comprehending the issue. We report only descriptive stats due to the qualitative nature of the data.

The name constraints seem to be less understood compared to the other cases: Firstly, the codes describing the problem are less frequent (only about half of the participants having at least one of these codes compared to about 80\% for other cases). Secondly, there are frequent codes explicitly admitting not understanding the issue (\textsc{NotKnow, NoInfo}) or misunderstanding (\textsc{Wrong}). The comprehension seems to have been better for the redesigned condition in some cases (see \cref{subsec:conditions-comprehension} for details).

The trust is compared in \cref{fig:trust-comparison}. As expected, there are again significant differences among the cases (Friedman ANOVA, $\chi^2(4)=183.5$, $p<0.001$). Pairwise comparisons (Dunn-Bonferroni) show that the certificate in the OK case was trusted significantly more ($median\;\;6$) and the one in the hostname mismatch case was trusted significantly less ($median\;\;0$) than other cases ($p<0.005$ for all comparisons). The expired, self-signed and name constraints cases did not significantly differ from each other ($median\;\;3$ for expired, $median\;\;2$ for self-signed and name constraints). The effect size was the largest for the comparison OK/hostname mismatch ($r=0.26$), followed by the comparison of OK with other cases ($0.15 \leq r \leq 0.18$).

Neither the high trust in the OK case nor the low trust in the hostname mismatch case is surprising. However, we see the self-signed and name constraints cases as over-trusted (at least when compared to the expired case): The expired certificate provided full authenticity assurances before expiration, but the self-signed certificate never did (indeed anyone could have created such a certificate). The name constraints case suggests malicious activity at the authority level, which is far more severe (the intermediate CA was prohibited from issuing the certificate, yet it did). Obtained trust assessments signal potential misunderstanding of security implications in these cases.

\subsection{Influence of the Previous Experience}
\label{subsec:results-previous-experience}

In this section, we investigate the influence of previous experience on trust ratings and case comprehension. Ordinal logistic regression was run to determine the effects of previous experience on trust ratings for all five certificate cases. Tested possible predictors included the number of years employed in IT, the number of previously used tools (possible covariates), formal education in computer science, self-reported knowledge of computer security/certificates and previous OpenSSL usage (possible factors). The model did not predict the trust rating significantly better than the intercept-only model in any of the cases ($p_{name.mism}=0.52$, $p_{self.signed}=0.09$, $p_{expired}=0.91$, $p_{name.constr}=0.58$, $p_{ok}=0.16$) In summary, no significant predictors of trust ratings were found.

We further wanted to inspect the influence of previous experience on case comprehension as defined by the comprehension codes (see \cref{subsec:results-opinions-comparing}). However, only hostname mismatch and name constraints cases had a sufficient number of participants not coded by any comprehension codes to allow for such analysis. Differences were investigated by the Mann-Whitney U tests for the number of years employed in IT, self-reported computer security and certificate knowledge and the number of previously used tools. For the hostname mismatch case, no significant differences were found. For the name constraints case, differences were significant only for the number of tools previously used ($U = 766, z = 2.913, p = 0.004$) with $mean\;2.54 \pm 1.04$ for those coded by at least one comprehension code and $1.77 \pm 1.16$ for those having none.

In summary, taking the self-reported previous experience into account, we find only very little influence on trust assessment or case comprehension. This result is in contrast to the 2017 study by Acar et al.~\cite{acar2017-github} that reports a significant difference in both functionality and security of programming tasks, depending on self-reported years of experience. However, a similar study by Naiakshina et al.~\cite{naiakshina2017developers} failed to find such a correlation.

\section{Analysis of used resources}
\label{sec:results-resources}

This section concentrates on the behavior of participants. We compare time spent on the individual cases and discuss what offline and online resources the participants had used.

\subsection{Task Times}
\label{subsec:resources-times}

\begin{figure}
    \centering
    \includegraphics[width=\textwidth]{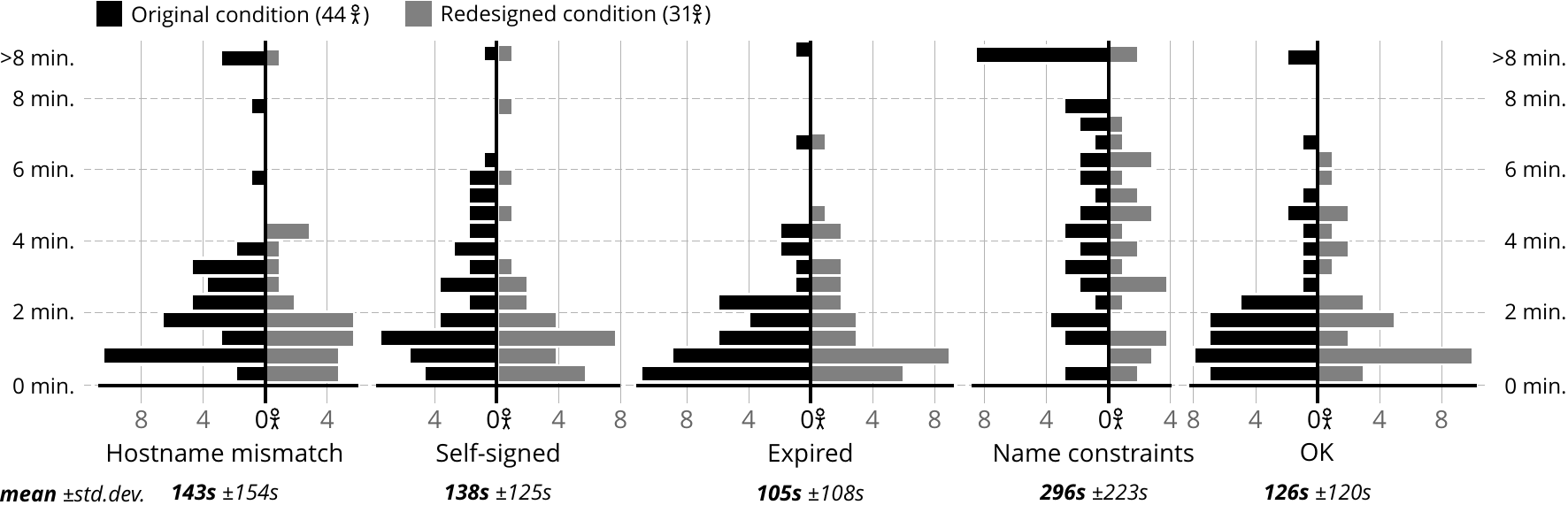}
    \caption{Histogram comparison of time spent on each certificate case, split by experimental condition. The vertical axis has task time divided into 30-second intervals and the horizontal axis shows the number of participants with time in the particular interval (all times over 8 minutes are merged for simplicity).     Note that the number of participants is different for the two conditions.}
    \label{fig:time-comparison}
\end{figure}

We have measured the time spent on each certificate case from the moment of seeing the validation message for the first time till the message of the following case was invoked
(the participants were instructed to work on cases in series). This time, therefore, includes the reading of the validation message, participant's investigation into the matter (e.g., an inspection of the certificate in question, reading manual pages, browsing online) and filling the trust evaluation into the provided form. Since reading the message and filling in the trust rating takes roughly the same time for all cases, the timing information mostly reflects the time spent on certificate flaw investigation. 

The histograms of task times for all the cases are plotted in \cref{fig:time-comparison}. It can be seen that the histograms for the name constraints case are noticeably flatter and there are far more participants with the extremely high times (over 8 minutes).
The differences among cases are significant (Friedman ANOVA, $\chi^2(4)=48.81$, $p<0.001$). Post-hoc analysis (Dunn-Bonferroni) shows a significantly higher mean time for the name constraints case ($mean\;\;4.9\;min.$, $median\;\;4.5\;min.$) when compared to other cases ($p \leq 0.001$ for all comparisons). All other cases had similar means of about 2 minutes (OK $2.1\;min.$, expired $1.75\;min.$, self-signed $2.3\;min.$, hostname mismatch $2.6\;min.$). The standard deviations for all cases are rather large (often comparable to the means), indicating the timing data might be less conclusive. Nevertheless, the comparison confirms the presence of obstacles in the comprehension process of the name constraints case, which is in alignment with the result from \cref{sec:results-opinions}.

Lastly, we investigated the correlation of previous experience and task times using partial Pearson’s correlation controlling for the experiment condition. There was a small but significant negative correlation in six out of 25 pairwise combinations: The task time in the expired case correlates with the years of IT employment ($r_{partial}(72)=-0.24$, $p<0.04$), university education ($r_{partial}(72)=-0.31$, $p<0.01$) and self-reported knowledge of computer security ($r_{partial}(72)=-0.28$, $p<0.02$). Task times in the self-signed case was correlated with the number of tools previously used ($r_{partial}(72)=-0.27$, $p<0.02$). Times in the name constraints case were correlated with the self-reported knowledge of computer security ($r_{partial}(72)=-0.25$, $p<0.04$) and in the OK case with the years of IT employment ($r_{partial}(72)=-0.27$, $p<0.02$). Overall, there seems to be a small but not systematic trend towards lower task times with increased previous knowledge (even though the trust rating could not be predicted from previous experience, see \cref{subsec:results-previous-experience}).

\subsection{Offline Sources}
\label{subsec:resources-offline}

During the task completion, participants had the opportunity to use several offline sources: 1) to inspect the \enquote{downloaded} certificates (display and manually assess their contents), 2) run local verification on them or 3) read local manual pages. Only a few people performed local verification using the OpenSSL's \texttt{verify} command (7\%, \n{5}) or read local manual pages regarding the error codes (7\%, \n{5}). However, 80\% of all the participants (\n{60}) inspected at least one \enquote{downloaded} certificate. The high rate was expected since the error message in neither condition was detailed enough (e.g., the message did not state the actual mismatched hostname) and the task specification explicitly offered a command to do the inspection. This supports the notion that the cases have essential additional information that could be stated directly in the error message to speed up the trust evaluation. Nevertheless, it says that a fifth of the participants made their trustworthiness assessment without this additional information (the length of the expiry, the mismatched hostname or the name constraint being violated).

\begin{figure}
    \centering
    \includegraphics[width=\textwidth]{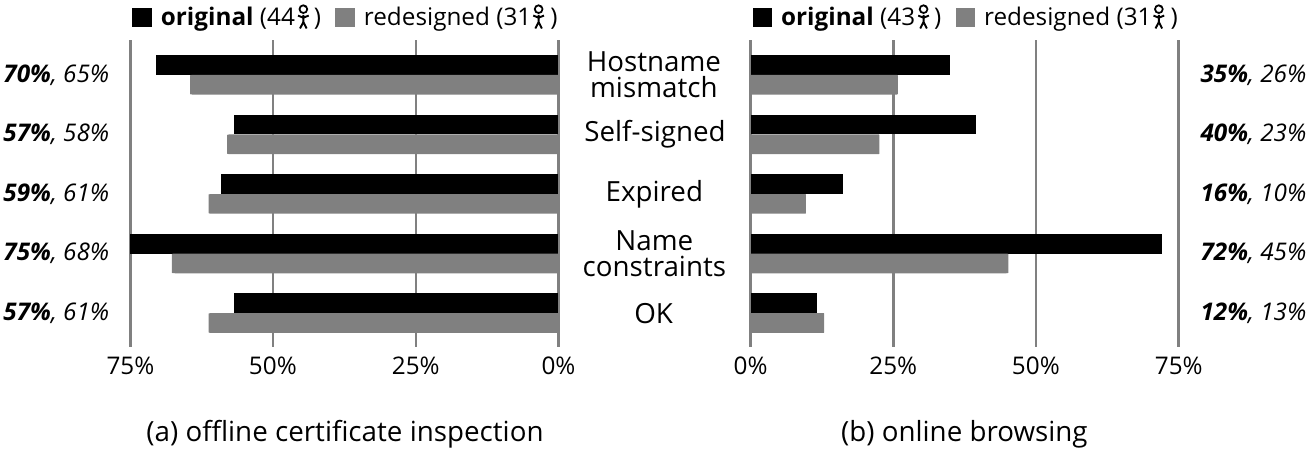}
    \caption{Overview of resource use per case (normalized by condition). Subfigure (a) displays offline certificate inspection during task completion. Subfigure (b) provides an overview of online browsing (one participant with connection problems is omitted).}
    \label{fig:resource-use}
\end{figure}

We present the overview of the offline certificate inspection, split per case, in \cref{fig:resource-use}, subfigure (a). Individual cases differ -- the name constraints case certificate was inspected the most (72\% of all the participants, \n{54}), followed by the certificate in the hostname mismatch case (68\%, \n{51}). The differences among the individual cases are significant (Cochran's Q test, $\chi^2(4)=13.91$, $p=0.008$). The pairwise analysis (Dunn-Bonferroni) shows that the name constraints case certificate was inspected by significantly more participants than the self-signed case ($p=0.027$). This result hints that, for some cases, additional information is more important than for others.

\subsection{Online Sources}
\label{subsec:resources-online}

Overall, about three-quarters of the participants (73\%, \n{55}) browsed the Internet. One of the participants experienced technical problems with the Internet connection -- we thus only report data for 74 participants.

85\% of those browsing online (\n{47}) used an online search engine to find the necessary information. Note that online search was considerably less used in the redesigned condition (for details, see \cref{subsec:conditions-times-resources}). Of all the participants doing an online search, 81\% (\n{38}) used the literal error text code from the validation message as the search query, 32\% (\n{15}) used the full verbatim error message and 66\% (\n{31}) had search queries with their own formulations. Having the error code (and even better the whole error message) present in the online documentation is, therefore, an essential strategy to have the documentation found when needed.

Next, we investigate the use of online resources per case. Based on the contents, we manually assigned websites to the experimental cases to which they are relevant. If the source contained the information relevant to multiple cases (e.g., commands to display the certificate), we categorized them as not uniquely belonging to any case. This manual attribution process gives only rough estimates since we were only able to assign the unique case to 61\% of the visited websites. Nevertheless, the analysis of these websites still gives valuable insights into the ecosystem.

The basic overview of the online resource use per case can be seen in \cref{fig:resource-use}, subfigure (b). The differences among the cases are significant (Cochran's Q test, $\chi^2(4)=68.85$, $p<0.001$) with the most participants browsing for information about the name constraints case (61\% of all the participants, \n{45}), less about self-signed and hostname mismatch cases (32\%, \n{24} and 31\%, \n{23}) and the least about the expired and OK cases (14\%, \n{10} and 12\%, \n{9}). The post-hoc analysis (Dunn-Bonferroni) shows significant pairwise comparisons for all combinations with the name constraints case ($p \leq 0.001$) and also among three other pairs ($p \leq 0.05$).
The inspection of the total number of websites ($n=136$, excluding search pages) shows that the most visited websites concerned the name constraints case ($n=62$, 45\%), followed by self-signed and hostname mismatch cases ($n=32$, 24\%; $n=30$, 22\%) and expired and OK cases ($n=8$, 6\%; $n=4$, 3\%). Both these analyses work as yet another indirect suggestion at comprehension obstacles for the name constraints case (the case is either unknown, complicated or the resources are poor, making people visit many more sites than for other cases).

Looking at used online resources in more detail, we can see that some domains were visited much more often than others. For an overview of the most visited domains, refer to \cref{tab:visited-domains}. The list starts with \textit{google.com} (85\% of the participants browsing online, \n{47}) and contains a mixture of vendor websites and unofficial documentation sources. The second most visited was the official documentation on \textit{openssl.org} (62\%, \n{34}), followed by multiple unofficial sources. Note that the website x509errors.cz was created by us for the condition with redesigned errors (for details on its usage, refer to \cref{subsec:conditions-times-resources}). Compared to previous work on developer resource use by Acar et al.~\cite{acar2016you}, we confirm the prevalent use of informal sources, but we still see vendor documentation as much used.

\begin{table}
\caption{The most visited domains for \n{55} browsing online, \n{34} in the original condition (\textsc{o.}), \n{21} in redesigned (\textsc{r.}).}
\rowcolors{2}{gray!15}{white}
\renewcommand{\arraystretch}{1.2}
\setlength{\tabcolsep}{0.5em}
\begin{tabular}{rrrlm{2.4cm}}
\rowcolor{white}
\% & \personSymbol & \textsc{(o. + r.)} & \textsc{Domain} & \textsc{Site type} \\ \hline
85\% & \n{47} & (34+13) & \href{https://google.com}{google.com} & search engine \\
62\% & \n{34} & (33 + 1)  & \href{https://openssl.org}{openssl.org} & documentation \\
29\% & \n{16} & (9 + 7)   & \href{https://sysadmins.lv}{sysadmins.lv} & personal blog \\
27\% & \n{15} & (0 + 15)  & \href{https://x509errors.cz}{x509errors.cz} & documentation \\
20\% & \n{11} & (8 + 3)   & \makecell[l]{
                            \href{https://stackoverflow.com}{stackoverflow.com}\\
                            \href{https://stackexchange.com}{stackexchange.com}} & Q/A forums by \newline Stack Exchange \\
16\% & \n{9}  & (5 + 4)   & \href{https://microsoft.com}{microsoft.com} & documentation \\
15\% & \n{8}  & (5 + 3)   & \href{https://github.com}{github.com} & code repositories \\
11\% & \n{6}  & (4 + 2)   & \href{https://mozilla.org}{mozilla.org} & documentation \\
11\% & \n{6}  & (4 + 2)   & \href{https://sslshopper.com}{sslshopper.com} & tutorial site
\end{tabular}
\renewcommand{\arraystretch}{1}
\label{tab:visited-domains}
\end{table}

\section{Influence of Redesigned Errors}
\label{sec:results-comparison}

In this section, we examine whether error comprehension by IT professionals can be influenced by a content change in the error messages and documentation (without a major redesign). The section reiterates over the presented results again, focusing on the observed differences between experimental conditions -- \n{44} for the original error messages and documentation versus \n{31} for the redesigned condition (\n{39}/\n{28} for interview codes).

\subsection{Case Comprehension and Trust}
\label{subsec:conditions-comprehension}

Let us look again at the most common interview codes. For the comprehension codes (see \cref{fig:understanding-comparison}): In the name constraints and expired cases, the comprehension codes occur notably more frequently in the redesigned condition (\textsc{Constraint} 31\% vs. 46\% in the redesigned condition, \textsc{CAProblem} 5\%/29\%, \textsc{CAConstr} 5\%/25\%, \textsc{NoLonger} 87\%/100\%, \textsc{OKBefore} 10\%/36\%). 
The redesigned condition further features a decrease in the codes that describe incomprehension (\textsc{Wrong} 31\%/25\%, \textsc{NotKnow} 28\%/11\%, \textsc{NoInfo} 13\%/7\%, absolute numbers in \cref{tab:interview-codes,tab:interview-codes2}).

Looking at other codes: In the OK case, more people did manual certificate checks in the redesigned condition (\textsc{ExtraCheck} 15\%/29\%), probably due to the new documentation suggesting to the user to perform extra checks. In the redesigned condition, there was an increase in believing that the self-signed and name constraints errors are consequences of an attack (\textsc{Attack}, 5\%/21\% for self-signed, 8\%/25\% for name constraints). 
Other than the few mentioned differences, the occurring codes are consistent between the conditions.

Let us now look back at the trust assessment overview in \cref{fig:trust-comparison}. We compared the differences between conditions using a Mann-Whitney U test for each case. The results indicate that, in the redesigned condition, the trust was significantly lower for the self-signed case ($U=433.00$; $p=0.006$, $z=-2.77$, $r=-0.32$). The mean trust decreased from $2.32\;\pm\;1.54$ ($median\;\;2$) to $1.35 \pm 1.58$ ($median\;\;1$). There was a similar trend for hostname mismatch and name constraints, but the differences are not statistically significant. 

All in all, the redesigned documentation seems to be no worse than the original one. It seems to significantly increase the comprehension of the flaw in the name constraints case and better conveys the possible attack vectors in the OK, self-signed and name constraints cases.
As we considered the self-signed case to be over-trusted, we see the influence of the redesigned documentation on the trust ratings as helpful, shifting the perception of IT professionals in the desired direction.

\subsection{Task Times and Used Resources}
\label{subsec:conditions-times-resources}

The overview of the task times across cases was given in \cref{subsec:resources-times}. For all cases, there was a general trend towards lower tasks times for participants in the redesigned condition (see \cref{fig:time-comparison}). In addition, there were notably fewer people with very high times (over 8 minutes) in the name constraints case (20\%, \n{9} for the original condition, 6\% \n{2} for the redesigned). Nevertheless, the differences were not statistically significant (Mann-Whitney U test for each case).

As for the resources used, roughly the same proportion of participants conducted the offline certificate inspection in both conditions (82\%, \n{36} for the original condition, 77\%, \n{24} for the redesigned). Similarly, the differences in online browsing were also insignificant (original: 79\%, \n{34}, redesigned: 68\%, \n{21}). Inspecting the differences by case (\cref{fig:resource-use}), there was a significant decrease in online browsing in the name constraints case for the redesigned condition (72\% to 45\%, Fisher's exact test, $p=0.029$). There was also a difference in the online search behavior -- significantly fewer people used online search in the redesigned condition (only 62\% of those who browsed online, while it was 100\% in the original condition, $\chi^2(1)=10.72$, $p=0.001$). Both these results support the hypothesis that the redesigned error messages and documentation positively influence resource use. Participants in the redesigned condition had to spend less effort to achieve similar trust assessment and comprehension for the expired, hostname mismatch and OK case. Furthermore, they even achieved better results for the self-signed case (trust assessment) and the name constraints case (comprehension). 

As detailed in \cref{subsec:methodology-conditions}, there was a link to the new documentation website x509errors.cz embedded directly in the re-worded validation messages. The link in the message was followed by 71\% of the participants in the redesigned condition who browsed online (\n{15}). It is important to mention that all the participants browsing online who did not use online search (\n{8}) visited only the linked documentation page. People opening x509errors.cz usually did so for multiple cases ($mean\;\;4.2 \pm 1.27$, $median\;\;5$). Everybody consulting the page at least once did so for the name constraints case. Incorporating a documentation link directly into the error message thus turned out to be highly influential -- the page was visited by the majority of browsing participants, often in multiple cases.

\section{Related Work}
\label{sec:related-work}

In this section, we give overview of the research on the causes of certificate validation errors and perceptions of the arisen TLS errors by end users and IT professionals. Furthermore, we briefly discuss influence of documentation and error messages on system security in general and the specific situation with documentation related to certificate validation.

\subsection{Root Causes of Certificate Validation Errors}
\label{subsec:related-work-root-causes}

To validate without errors, a TLS certificate needs to be within a specified validity period, it must not be revoked, server hostname needs to match the subject name and it needs to have a valid chain-of-trust up to a trusted root authority. If any of these requirements are not met, the identity of the certificate provider cannot be determined. While certificate validation errors indicate a problem with the certificate, the problem source is often of non-malicious nature. 

Akhawe et al.~\cite{akhawe2013here} monitored almost 4 billion TLS handshakes generated by 300\,000 browser users throughout nine months in 2012--2013. As attacks are rare, Akhawe et al. concluded that most errors were caused by a misconfiguration on the server side. They found that 1.54\% of the observed certificates do not validate. The most prevalent cause is an unknown issuer (about 70\%), followed by name validation errors (about 18\%), expired certificates (about 7\%), and self-signed certificates (about 3\%). Within the errors found, they identify low-risks scenarios that they suggest to mitigate with more relaxed validation policies. For example, 25\% of websites with expired certificates were accessed a maximum of seven days after the expiration date, indicating that the problem was fixed within one week. Similarly, due to the strict name matching policy, more than 50\% of name validation errors had a matching second-level domain. While the adoption of more relaxed validation policies could help avoid bothering users with false warnings~\cite{akhawe2013here}, it could also introduce some false negatives potentially concealing an attack.

In a large-scale study on browser certificate warnings from 2016, Acer et al.~\cite{acer2017-https-errors} found that server errors accounted for less than half of all browser warnings. The rest were caused by network or client-side errors such as misconfigured client clocks causing the expiration errors. They presented mitigations to prevent network and client-side errors and subsequently succeeded to decrease the error rate in the patched browser by 25\%. While this result is encouraging, server errors remain a significant source of false warnings. Currently, it is mostly left to the end user to decide whether the TLS warning is a false positive or not.

\subsection{End Users and TLS Certificates}
\label{subsec:related-work-end-users}

In contrast to IT professionals, the perceptions of end users with respect to TLS warnings have been thoroughly investigated. Studies indicate that a considerable portion of end users does not understand the causes of certificate warnings. Sunshine et al.~\cite{sunshine2009crying} surveyed 400 browser users in 2008 to investigate reactions and understanding of TLS warnings (expired, unknown issuer, hostname mismatch). They found that the majority of users did not understand the warnings. 
Interestingly, those users who understood the warning chose more often to adhere to hostname mismatch warnings but considered warnings for expired certificates as low risk (a similar trend is also present in our study in trust assessment, see \cref{subsec:results-opinions-comparing}). 

A study by Felt et al.~\cite{felt2015improving} investigated whether users can understand the threats they might be facing when seeing TLS warnings. The most users misidentified the threat source (more than 50\% thought it was malware) and less than 20\% understood what data was actually at risk (most overestimated the risk).

Stojmenovic et al.~\cite{stojmenovic2018building} tested an intervention to help users build a better mental model of TLS certificates. While they succeeded to improve trust in the website in a benign scenario with extended validation certificates, their interventions did not have an effect in an attack scenario with a fake website deployed and a domain-validated certificate. 

Reeder et al.~\cite{reeder2018experience} conducted a study on browser warnings to find out why users do not adhere to warnings. The most common reasons to ignore a TLS warning was connecting to a known trusted website, e.g., company-internal or their own (both opinions also arose in our study, see \cref{subsec:self-signed}). Whereas warning adherence rates increased over time for major browsers, comprehension rates remain low and misconceptions are still a problem~\cite{felt2015improving}. 

In summary, research shows that end users seem to lack the understanding of certificate warnings and their security implications. In most cases, IT professionals could set the appropriate decision already during system development. However, to do so, it is crucial that they understand the errors and security consequences themselves -- which is why we decided to focus on IT professionals.

\subsection{IT Professionals and TLS Certificates}

The environment of X.509 certificates and TLS infrastructure is rather complicated, allowing for a wide variety of things that can go wrong~\cite{clark2013-sok-ssl}.
As shown by Krombholz et al.~\cite{krombholz2017have}, TLS seems to be complicated to set up and configure even for IT professionals. In a conducted usability study on the deployment of TLS on Apache servers, even people with network security knowledge struggled to configure the server correctly. When being asked about the usability shortcomings of the deployment process, participants noted the unavailability of best practice tutorials, misleading terminology and error messages, as well as a weak default configuration. 

Apart from the server configuration, there are deficiencies in programming APIs as well. Georgiev et al.~\cite{georgiev2012most} showed that options, parameters, and return values of widely used TLS implementations can be easily misunderstood by developers, often leading to certificate validation in non-browser software done incorrectly or not at all.

Ukrop and Matyas present another usability study~\cite{2018-rsa-ukrop}, in which participants were to generate and validate certificates with the OpenSSL command line interface. Almost half of the participants in the experiment assumed that they had succeeded in creating a valid self-signed certificate, although they had not. Only a fraction (19\%) of participants were able to correctly validate the provided certificates, suggesting the usability of OpenSSL is also far from optimal.

Usability issues, as described above, are a major obstacle in getting TLS configurations and certificate handling right for developers, testers and administrators alike. However, security issues are not only the result of poor usability -- they may be caused by institutional or organizational factors out of control of the administrators~\cite{dietrich2018investigating}. In the context of TLS certificates, Fahl et al.~\cite{fahl2014eve} found that almost two-thirds of the non-validating certificates were deployed deliberately. One reason included the websites were never supposed to be publicly accessible (this manifested itself also in our study, see \cref{subsec:self-signed}). Another reason for deploying self-signed certificates was to save money (also mentioned in our interviews, but only marginally).

However, even being able to deploy TLS correctly does not tell us much about IT professionals' perception of certificate flaws they encounter when connecting to other servers. In this respect, our work complements the previous research mentioned above.

\subsection{Documentation and Error Messages}

Our attempts to improve the understanding of security consequences of certificate validation errors were based on re-worded error messages and redesigned documentation.
Bralo-Lillo et al.~\cite{your-attention-please} found that when end-user security dialogues get appropriately adjusted, significantly fewer people ignore clues indicating they may be at risk.

The importance of usable documentation is discussed by Georgiev et al.~\cite{georgiev2012most}, stressing the importance of clean and consistent error reporting. Error handling is also seen as vital by Gorski and Lo Iacono \cite{gorski2016towards}. In further work~\cite{new-api-warnings}, they prototyped an API-integrated security advice system successfully nudging developers towards secure decisions.
However, as their novel approach is less compatible with the current methods, we decided to do only smaller content changes in our study. Compared to our work (updating just the content of the errors and accompanying documentation), Gorski et al.~\cite{new-api-warnings} proposed and tested a significant redesign of the error-reporting system to nudge towards secure behavior among IT professionals.

Moreover, several works emphasized the crucial role of documentation for security API usability. Robillard argues~\cite{robillard2009makes} that documentation and code examples are an essential information source for developers to learn how to use an API. Concerning documentation in the security context, Acar et al.~\cite{acar2016you} have shown that IT professionals resort to online forums and tutorials on the Internet, often helping them get the code functional but not necessarily secure. Therefore, providing easily accessible documentation with security information and examples of secure code is of crucial importance~\cite{acar2017comparing, acar2016you}.

\subsection{Documentation of Certificate Validation Errors}
\label{subsec:related-work-cert-errors}

Certificate validation is only a small part of the security operations provided by cryptographic libraries. According to the study of publicly available RSA keys by Nemec et al.~\cite{nemec2017measuring}, the most common library (by far) seems to be OpenSSL, with GnuTLS, OpenJDK, Botan, mbedTLS, WolfSSL, Libgcrypt and Microsoft CryptoAPI also being reasonably used. Inspection of certificate validation errors in these libraries shows that almost every library has its own set of errors. The notable exception is WolfSSL which reuses a subset of OpenSSL errors and refers to the OpenSSL manual in its documentation~\cite{wolfssl-docs-refer-openssl}.

A more detailed investigation of certificate validation error messages and documentation for OpenSSL, GnuTLS, Botan, mbedTLS and Microsoft CryptoAPI is summarized in \cref{tab:documentation-overview}. The number of errors concerning certificate validation varies a lot -- from only 17 (GnuTLS) to as high as 78 (OpenSSL). However, looking at the error messages (human-readable strings defined in code) and documentation, almost all these libraries share the brevity of explanation. While the shortness of the error messages is not surprising, the median documentation of the error is also only 8--16 words.

\begin{table}
    \centering
    \caption{Statistics of messages/documentation corresponding to certificate validation errors in multiple cryptographic libraries.}
    \rowcolors{2}{white}{gray!20}
    \renewcommand{\arraystretch}{1.2}
    \begin{tabular}{lc|>{\centering}p{1.4cm}>{\centering}p{1.4cm}|>{\centering}p{1.4cm}>{\centering}p{1.4cm}|p{4.4cm}}
        & & \multicolumn{2}{c|}{\parbox{2.9cm}{\centering \textsc{Error message\\ length (words)}}} & \multicolumn{2}{c|}{\parbox{2.9cm}{\centering \textsc{Error docs.\\ length (words)}}} & \multirow{2}{*}{\textsc{Library URL}} \\
        \rowcolor{white}
        \multirow{-2}{*}{\textsc{Library}} & \multirow{-2}{1.4cm}{\centering\textsc{Num. errors}} & \textsc{avg.} & \textsc{med.} & \textsc{avg.} & \textsc{med.} & \\ \hline
        OpenSSL & 78 & 4.7 & 5 & 11.2 & 9 & \href{https://www.openssl.org/}{www.openssl.org} \\
        GnuTLS & 17 & 7.8 & 7 & 12.8 & 9 & \href{https://www.gnutls.org/}{www.gnutls.org} \\
        Botan & 50 & 4.7 & 4 & -- & -- & \href{https://botan.randombit.net/}{botan.randombit.net} \\
        mbedTLS & 20 & 8.5 & 8 & 8.5 & 8 & \href{https://tls.mbed.org/}{tls.mbed.org} \\
        MS Crypto API & 58 & -- & -- & 19.3 & 16 & \multicolumn{1}{m{4.4cm}}{\href{https://docs.microsoft.com/en-gb/windows/win32/seccrypto/cryptography-portal}{docs.microsoft.com/windows/ \hspace*{1em}/win32/seccrypto}} \\
    \end{tabular}
    \label{tab:documentation-overview}
\end{table}

An initiative exists at \href{https://x509errors.org}{x509errors.org}~\cite{x509errors-org}, aiming to consolidate the errors and their documentation across multiple libraries. The project's goal is to create a mapping of the error systems of different libraries along with their documentation and example certificates consolidated in one place. However, the overview could be improved by incorporating results of studies on user understanding (such as this one) or occurrence statistics (such as the study by Akhawe et al.~\cite{akhawe2013here}).

\section{Studies with IT professionals: Lessons Learned}
\label{sec:admin-studies}

As usable security studies involving IT professionals are still infrequent, this section discusses our \enquote{lessons learned} regarding conducting studies with IT professionals (as opposed to studies with end users). The experience stems mostly from the presented study, but also incorporates the experience from a similar previous study~\cite{2018-rsa-ukrop}. Conclusions are compared to other similar summaries~\cite{2019-ccs-tiefenau,naiakshina2017developers,robust-experiment-design}.

\subsection{Participant Sampling}

Studies with IT professionals are often done on students~\cite{krombholz2017have} or convenience samples, such as GitHub users~\cite{acar2017-github}. Although studies suggest research with this sampling may be generalizable in some cases~\cite{replicating-ssl-warnings-study, acar2017-github}, such approximations are not ideal.

As mentioned in \cref{subsec:methodology-setting-recruitment}, we have sampled participants from the attendees of a large IT conference, having set up a \enquote{research booth\punct{.}} Even with the task spanning 30--40 minutes, we were easily able to recruit participants. Many of them joined with enthusiasm, enquired about the research afterward and wanted to be informed about the results. Although some sort of compensation seems essential, simple conference merchandise worked sufficiently. Multiple participants approached us having seen mugs or beanies other participants got -- thus, we recommend the provided merchandise to be unique to your booth. Furthermore, a poster with IT comic strips worked as a great, yet simple, attention grabber. Such an attention grabber makes random passers-by stop at the booth and thus be approachable to join the research. The increased chances to grab all passers-by also helps reduce the self-selection bias of the sample.

The participant sample created this way can be quite heterogeneous with respect to their previous experience. Although diverse previous experience is common~\cite{2018-so-dev-survey} and thus ecologically valid, some assessment is needed -- different skill set can~\cite{acar2017-github} but does not have to be~\cite{naiakshina2017developers} a great confounding factor. We agree with Tiefenau et al.~\cite{2019-ccs-tiefenau} that more work on reliably assessing IT skills is needed for reliable pre-screening or participant sample analysis.

In summary, we see conducting experiments at conferences and similar venues as a cost-effective alternative to online studies in controlled environments~\cite{developer-observatory}. We suggest not to underestimate the participant sample analysis and focus on previous experience in the field.

\subsection{Educational Debriefing}

As briefly mentioned in \cref{subsec:methodology-task}, our study design contained an educational debriefing. After the participant elaborated on their opinions and reasoning in the interview, the experimenter explained the unknown cases and corrected potential misconceptions. The participants were very appreciative of this and often had additional questions. Thus, we would suggest including educational content in the experiments for professional users if time and topic permit.

\subsection{Measurement Automation}

To preserve the ecological validity of the studies, researchers often leave tasks open (without constraining the participants). As compared to end users, tasks intended for IT professionals tend to have a longer duration and more degrees of freedom, often resulting in many non-linear ways to solve them.

Measuring such tasks comes with multiple issues, as the think-aloud protocol is less applicable due to the length and complexity~\cite{naiakshina2017developers} and less constrained solutions are challenging to measure. Such experiments, therefore, benefit from a detailed analysis of needed measurements before the experiments and during pilot testing. Most of them should then be measured and collected automatically during the task, as also observed by Tiefenau et al.~\cite{2019-ccs-tiefenau} (common examples include terminal records, browser histories and action timings).

Our recommendation is to still keep a low-level unifying fail-safe measurement (usually a screen or video recording) for manual corrections. However, one should not rely on it entirely as it may be extremely resource-intensive to code.

\subsection{Focus on Observed Behaviors}

The study of Wash et al.~\cite{self-reported-security} suggests that security research based on self-report is only reliable for a small subset of behaviors. Although the study was conducted on college students, the same seems to hold for IT professionals as well. An example from the presented study: Only 65\% of the participants who self-reported that the expiry duration makes a difference to their perceived trust actually inspected the certificate (\n{36/55}). This result hints at a possible disparity between self-reported and observed actions.

A similar experience was reported by Sotirakopoulos et al.~\cite{replicating-ssl-warnings-study}, observing a substantial disparity between the actions their participants claimed they would take and the actual actions during the laboratory tasks. Thus, for future studies, we recommend focusing on measurable observed behaviors or a combination of self-reported and observed behaviors.

\subsection{Counterbalancing Answer Scales in Questionnaires}

\begin{figure}
    \centering
    \includegraphics[width=\textwidth]{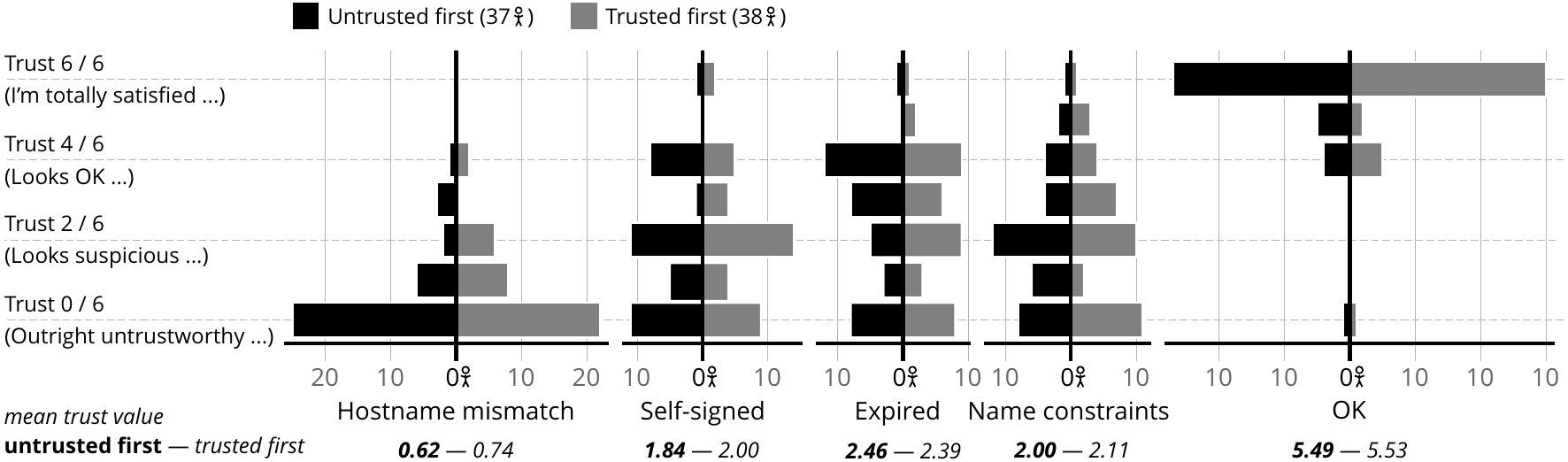}
    \caption{Investigating the effect of counterbalancing Likert scale direction in trust evaluation per case.}
    \label{fig:reversed-likert-trust}
\end{figure}

Questionnaire methodology research~\cite{likert-direction-yes} shows the response order in Likert scales may bias the participants towards the first acceptable option (the primacy effect), thus biasing the results. However, this effect is not conclusively reproduced~\cite{likert-direction-no}.

As detailed in \cref{sec:methodology}, our study used reversed direction on Likert scales for half of the participants (trust scales, questions on previous knowledge, SeBIS questions). As can be seen in \cref{fig:reversed-likert-trust}, no significant differences were observed for the trust scales (Mann-Whitney U test). Neither were significant differences observed for the pre-task questionnaire questions (Mann-Whitney U tests).

Overall, the order of Likert scales in our study seems not to affect the results. Nevertheless, we argue that if such counterbalancing takes only a little effort, future studies should do it as it strengthens the methodology (with it you can check if the order effects were present or not).

\section{Discussion and Conclusions}
\label{sec:conclusion}
We conducted a study with 75 people working in IT to understand how they perceive flawed TLS certificates. 
Even though similar studies had been conducted for end users before, this is the first such study for IT professionals. 
We further investigated how we can influence comprehension and resource use by re-wording the error messages and redesigning documentation.

\subsection{Trust Decisions Are Not Binary}

From a security point of view, the decision to trust a certificate deployed by someone else is binary: if the certificate does not validate, it should not be trusted. Our study results show that the trust decisions of IT professionals are far from binary and depend on the particular flaw and its context.

For example, for a majority of participants, the trust in expired certificates heavily depends on the time elapsed from the expiry date, indicating the expiration time may be used as a proxy to determine whether it was a misconfiguration or an attack. This was also evident in the qualitative results, where many participants mentioned that expired certificates are likely a mistake and rather common. This perception may be influenced by the actual reality of the Internet: 25\% of websites with expired certificates are accessed a maximum of seven days after the expiration date, indicating that the problem is often fixed within one week~\cite{akhawe2013here}. Moreover, several participants suggested that the reputation of the certificate subject also plays a role.

Besides the fuzzy trust in the expired certificate, IT professionals seem to overly trust the self-signed and the name constrained certificates: both were rated similarly to the expired case even though the security implications are rather worse. 
Furthermore, we have to keep in mind that this may be only a lower bound on the trust in the broader population due to a social desirability bias (see \cref{subsec:methodology-limitations}).
The trust evaluation of the name constraints case is especially worrying, as this case was also poorly understood (even by our slightly above-experienced sample).
For the self-signed case, the context was important: expecting a self-signed certificate on a (known) server or using it for internal or testing purposes is seen as a less severe problem.

\subsection{Security Implications}

The security implications of trusting flawed certificates are different in different cases.
For example, it may be acceptable for the system administrator to trust a flawed certificate deployed deliberately on their own server for testing purposes (the potential security consequences concern only themselves).
However, for IT professionals developing applications used by (potentially) millions of end users, the situation is different as their decision impacts all the end users.

Delegating the decision to the end users is generally a bad idea since they tend to make uninformed decisions due to the lack of understanding (see \cref{subsec:related-work-end-users}). Moreover, given that certificate flaws are common, but mostly benign, end users could further lose their trust in the security of the TLS ecosystem~\cite{sasse2015scaring,krombholz2019ifhttps}, In most cases, IT professionals should, therefore, make security decisions during the development, testing and deployment process as they are capable of better-informed decisions.

Even though IT professionals have a better potential to make appropriate security decisions than end users, our study shows some still fairly trust invalid certificates: 21\% of the participants (\n{16}) say the self-signed certificates \qsimple{looks OK} or better; for name constraints certificate it is 20\% (\n{15}). Thus, certificates with errors are still trusted by some of the IT professionals working with them.

While our work presents valuable insights into IT professionals' perceptions of flawed certificates, further research is needed to investigate the coping strategies people in IT use when encountering such certificates in the wild.

\subsection{Error and Documentation Design Matters}

We show that even a simple content redesign of the error messages and documentation matters: It positively influenced the comprehension of the errors and trust assessment of the certificates (see \cref{subsec:conditions-comprehension,subsec:conditions-times-resources}). Both over-trusted cases (self-signed and name constraints) were rated lower and the comprehension in the name constraints case improved considerably (expressed by more frequent comprehension codes, less frequent incomprehension codes, lower task times and less Internet browsing).

The majority of participants with the opportunity (71\%, \n{15}) followed the link provided in the re-worded error message, pointing to an excellent and cost-effective design opportunity to lead users to a trusted unified documentation source. 

\subsection{Methodological Considerations}

Based on our experience with usable security experiments on IT professionals, we summarize several study design suggestions. Firstly, IT conferences seem to be an excellent sampling opportunity for studies involving IT professionals. As such samples tend to be quite heterogeneous, controlling for previous experience is crucial. Secondly, the inclusion of educational debriefing after empirical experiments may be beneficial as IT professionals appear to be interested to learn. Thirdly, observed behaviors should be preferred to self-reported ones. As much data as possible should be collected automatically to ease the processing.

\subsection{Future Work}

While we obtained new insights into participants' perceptions of flawed TLS certificates, more work is needed to investigate what influences these perceptions and what coping behavior they cause.

Nonetheless, we should design systems embracing the complexity of trust decisions IT professionals make instead of forcing them to choose a binary option. We should pay more attention to error message and error documentation design as it turns out that even simple content changes can have significant effects.
The name constraints extension seems to be very poorly understood, suggesting its wider deployment might be problematic without extra educational efforts.
We plan to propose a few simple patches to OpenSSL (and possibly other libraries), re-wording name constraints error message and the accompanying documentation. More significant changes, such as error messages linking to good documentation or rewriting the documentation to clearly state the security implications, may also be beneficial but would require a discussion with the developer community.

In conclusion, the TLS certificate ecosystem proves to be complicated enough to produce a wide variety of attitudes, opinions and misconceptions even among IT professionals. To improve the situation, designers need to knowingly strive for good developer usability. 

\begin{acks}
We appreciate the support of Red Hat Czech and DevConf. We are particularly grateful to Nikos Mavrogiannopoulos, Mat\'{u}\v{s} Nemec and Tobias Fiebig for insightful comments, Heider Wahsheh for research help and to Vlasta \v{S}\v{t}avov\'{a}, Ag\'{a}ta Kru\v{z}\'{i}kov\'{a} and Martina Olliaro for their help with the experiment. We also thank all experiment participants.
\end{acks}

\bibliographystyle{ACM-Reference-Format}
\bibliography{main}

\appendix
\section{Task details}
\label{app:questionnaires}

This appendix contains the full formulations of the pre-task questionnaire, the task and its context (as given to the participant) and questions guiding the post-task semi-structured interview.

\subsection{Pre-task Questionnaire}

To better put your opinions in context, we need to know what previous experience you have.

\begin{enumerate}
    \item How many years have you been employed in the IT sector? (enter '0' if not employed) \textit{[number answer field]}
    \item Do you have a university degree in computer science? \textit{[No formal education; Bachelor degree; Master degree; Postgraduate degree]}
    \item How would you describe your knowledge of computer security in general? \textit{[Poor; Fair; Good; Very good; Excellent]}
    \item How would you describe your knowledge of X.509 certificates? \textit{[Poor; Fair; Good; Very good; Excellent]}
    \item Which of these tools/libraries have you ever used? Tick all that apply. \textit{[OpenSSL, GnuTLS, Network security services (NSS); Java Keytool; Windows Certutil; Let's Encrypt Certbot]}
    \item Have you participated in our last year’s experiment here at DevConf.CZ? \textit{[Yes; No]}
\end{enumerate}
These questions were followed by the full Security Behavior Intentions Scale~\cite{egelman2015scaling} with randomized question order.

\newpage
\subsection{Experiment Task}
\noindent\textsc{Context}
\begin{itemize}
    \item Imagine you want to improve DevConf registration system by allowing login with third-party accounts (GitHub, Fedora Project, Google, Microsoft, Facebook). You decide to contribute by coding it yourself and then submitting a pull request.
    \item Imagine, you have already created a small program testConnection.c testing the connection and validating server certificates. However, some server certificates seem to be problematic. Now, you should check them out.
\end{itemize}
\textsc{Your task}\\
Use the testing program \texttt{testConnection.c} to validate the certificates of the servers. That is, for each server:
\begin{enumerate}
    \item Run the testing program using \texttt{./testConnection <server\_name>}
    \item If the certificate is not valid, try to understand the problem and risks (read the documentation, tutorials, ...).
    \item Decide how much do you trust the server having this certificate? (Fill in the provided form.)
\end{enumerate}
\textsc{Details}
\begin{itemize}
    \item No coding or bug fixing is required from you, the program \texttt{test\-Connection.c} does the verification correctly (showing you all occurred errors).
    \item You (ultimately) trust the \textit{DevConf Root CA} certificate authority.
    \item You can view the downloaded certificates using\\\texttt{certtool -i --infile <chain.pem>}
    \item When trying to understand the problem, you can use any resources you want (manual pages, tutorials, Google, ...).
    \item Server order does not matter, but investigate the next server only after finishing the previous -- we measure individual times.
    \item The testing program \texttt{testConnection.c} is located in \texttt{\textasciitilde/Documents}, start your task by running it in the terminal with no arguments (displays help).
\end{itemize}

\begin{table}[t]
\centering
\caption{Certificate validation messages in the original and redesigned conditions.}
\label{tab:errors}
\renewcommand{\arraystretch}{1.3}
\begin{tabular}{>{\bf}m{1.75cm}|>{\raggedright}m{6.1cm}|>{\raggedright\arraybackslash}m{6.9cm}}
\textsc{Case} & \textsc{Original error message} & \textsc{Redesigned error message} \\[0.3em] \hline
Hostname mismatch & hostname mismatch \newline(X509\_V\_ERR\_HOSTNAME\_MISMATCH) & The server hostname does not match the certificate subject name. (X509\_ERR\_HOSTNAME \_MISMATCH, see http://x509errors.cz) \\
Self-signed & self signed certificate (X509\_V\_ERR\_\newline{}DEPTH\_ZERO\_SELF\_SIGNED\_CERT) & The certificate is self-signed and not found in the trust store. (X509\_ERR\_SELF\_SIGNED, see http://x509errors.cz) \\
Expired & certificate has expired \newline(X509\_V\_ERR\_CERT\_HAS\_EXPIRED) & The certificate has expired or is not yet valid. \newline(X509\_ERR\_EXPIRED, see http://x509errors.cz) \\
Name \newline constraints & permitted subtree violation \newline(X509\_V\_ERR\_PERMITTED\_VIOLATION) & The subject name violates constraints set by CA. \newline(X509\_ERR\_NAME\_CONSTRAINTS \_VIOLATION, see http://x509errors.cz) \\
OK & ok (X509\_V\_OK) & All performed checks passed. \newline(X509\_OK, see http://x509errors.cz) \\
\end{tabular}
\renewcommand{\arraystretch}{1}
\end{table}

\subsection{Trust Evaluation Scale}
\noindent
How much do you trust the server having this certificate?
\begin{itemize}
    \item Trust 6 / 6 (I'm totally satisfied. If it were my bank's website, I would log in without worries.)
    \item Trust 5 / 6
    \item Trust 4 / 6 (Looks OK. I would log in with my library account, but not with my bank account.)
    \item Trust 3 / 6
    \item Trust 2 / 6 (Looks suspicious. I will read the page, but I will not fill in any information.)
    \item Trust 1 / 6
    \item Trust 0 / 6 (Outright untrustworthy. It is not safe to browse or to trust any information there.)
\end{itemize}

\subsection{Post-task Interview}

Now please describe in your own words what was the problem with the certificates case by case. (Fedora project, Facebook, GitHub, Google, Microsoft)

\smallskip\noindent
\textit{Note: This open-ended question usually took over half of the interview time. The interviewers prompted the participant to elaborate further as necessary. Afterward, the following closed questions were posed if not answered spontaneously during the open-ended discussion.}

\begin{itemize}
    \item Did you check for how long is the certificate expired? \textit{[Yes; No]}
    \item Does it make a difference for you how long are they expired? \textit{[Yes; No]}
    \item If yes, how would you rate the certificate expired... \textit{[yesterday; a week ago; a month ago; a year ago]}
    \item Do you know what it means to revoke the certificate? Please explain.
    \item Did you know that expired certificates are no longer included in Certificate Revocation Lists? \textit{[Yes; No]}
    \item If not, does this change your trust rating? \textit{[Yes; No]}
    \item Which one was the most difficult to understand? Why? \textit{[Hostname mismatch; Self-signed; Expired; Name constraints; OK; None]}
    \item Can you name any examples of exceptionally well or exceptionally poorly done error messages you encountered in your life?
    \item Which of the following parts should it have? \textit{[Error number; Text code; Human-readable description]}
    \item Imagine you can have unlimited time and resources to change anything regarding error reporting and the developers of the world would comply. What would you want?
\end{itemize}

\section{Redesigned Documentation}
\label{app:documentation}

This appendix contains the full wording of the original and redesigned error messages (see \cref{tab:errors}) and the redesigned documentation that was available during the experiment at \href{http://x509errors.cz}{x509errors.cz}. A link to this website is the part of the redesigned error messages.

\setlist[description]{font=\normalfont\scshape}
\subsection{Hostname Mismatch Case}
The server hostname does not match the certificate subject name. (X509\_ERR\_HOSTNAME\_MISMATCH)
\begin{description}
    \item[Explanation.] The domain name provided by the server you are connecting to does not match the subject name of the certificate.
    \item[Security perspective.] Your communication will be encrypted, but you communicate with a different (maybe malicious) server from the one listed in the certificate. However, it can also be caused by malicious attackers pretending to be the server you are connecting to.
    \item[Next steps.] See the Common Name (CN) or the Subject Alternative Name extension (SAN) in the certificate and compare the value with the domain name of the server. In case of web servers, the error can be caused by improper redirect configuration between valid web aliases (e.g., the version of the site without the \enquote{www} in the domain name).
\end{description}

\subsection{Self-signed Case}
The certificate is self-signed and not found in the trust store. (X509\_ERR\_SELF\_SIGNED)
\begin{description}
    \item[Explanation.] A self-signed certificate is signed by the same entity whose identity it certifies. In technical terms, a self-signed certificate is one signed with its own private key. This is common for root keys (usually pre-installed on your system).
    \item[Security perspective.] No certificate authority issued the certificate. Anyone can create a certificate with these parameters. It means a really low level of security and trust. Since this self-signed certificate is not "managed" by a CA, there is no possible revocation.\\
    It can be trusted only if you control every machine between you and the server or you check that the keys in the certificate are what you expect them to be. However, it can also be caused by malicious attackers pretending to be the server you are connecting to.
    \item[Next steps.] If you can, try to validate the certificate in a different way – compare the fingerprint with what is expected, compare it with the certificate obtained on previous visits, etc.
\end{description}

\subsection{Expired Case}
The certificate has expired or is not yet valid. (X509\_ERR\_EXPIRED)
\begin{description}
    \item[Explanation.] Most of the certificates have a defined period of validity (the "not before" and "not after" validity fields). The expiration date is specified by the issuing certificate authority, usually for 1--2 years (there could be cases with shorter or longer validity).
    \item[Security perspective.] Your communication will be encrypted, but the identity of the server stated in the certificate is now no longer guaranteed by the trusted certificate authority.\\
    However, the certificate revocation status is no longer published. That is, the certificate might have been revoked long ago, but it will no longer be included in the CRL. That means you cannot know if the certificate is already revoked or not.
    \item[Next steps.] In some cases, the error may be caused by the misconfiguration on the client side -- make sure your local time settings are correct. If you are responsible for the certificate, contact the company that issued it to learn how to renew it. If not, inform the appropriate administrator.
\end{description}

\subsection{Name Constraints Case}
The subject name violates constraints set by CA. (X509\_ERR\_NAME\_CONSTRAINTS\_VIOLATION)
\begin{description}
    \item[Explanation.] By default, all certificate authorities can issue certificates for any subjects. One way to constrain this risk of issuing a malicious certificate is to limit CAs to issue only for certain names, using the "name constraints" extension for X.509 certificates. The occurred error says that the authority in the chain was prohibited from issuing the certificate for that particular subject.
    \item[Security perspective.] The certificate authority was not allowed to issue the certificate it issued. This suggests suspicious activity at the authority level.
    \item[Next steps.] See the Name Constraints extension in the certificate authority and compare the allowed/restricted name ranges with the Common Name (CN) or the Subject Alternative Name extension (SAN) in the issued certificate. Report the certificate to the issuing authority.
\end{description}

\subsection{OK Case}
All performed check passed. (X509\_OK)
\begin{description}
    \item[Explanation.] No errors occurred during the performed validation tasks. Make sure that everything you wanted to check was checked.
    \item[Security perspective.] Everything seems OK.
    \item[Next steps.] Check the hostname, the revocation status (by means of CRL or OCSP) check having appropriate verification flags.
\end{description}

\end{document}